\documentclass[preprint]{elsarticle}
\pdfoutput=1
\makeatletter
\def\ps@pprintTitle{%
     \let\@oddhead\@empty
     \let\@evenhead\@empty
     \def\@oddfoot{\footnotesize\itshape
       Accepted for publication in \ifx\@journal\@empty Elsevier
       \else\@journal\fi\hfill January 1, 2015}%
     \let\@evenfoot\@oddfoot}
\makeatother
     
\usepackage{amsopn}
\usepackage{amssymb}
\usepackage{amsmath}

\usepackage{booktabs,pbox}

\usepackage{subfigure} 
\usepackage{url,color}
\usepackage{xcolor}
\usepackage{algorithmicx}
\usepackage{algorithm}
\usepackage{algpseudocode}
\usepackage{bmpsize}
\usepackage{graphicx}

\newtheorem{definition}{Definition}[section]

\newcommand{\para}[1]{\vspace{0.5em}\noindent\textbf{#1}}
\newcommand{\ZKYC}{{{\em ZKYC}\cite{Zhang2012}}}
\renewcommand{\vector}[1]{\overrightarrow{#1}}
\newcommand{\tile}{\mathcal{T}}

\newcommand{\Q}{\mathcal{Q}}
\newcommand{\p}{\mathcal{P}}
\biboptions{square}
\journal{Information Processing \& Management}
\begin{document}
\begin{frontmatter}
\title{Geo-Temporal Distribution of Tag Terms for Event-Related Image Retrieval}

\author{Massimiliano Ruocco} 
\author{Heri Ramampiaro}

\address{Norwegian University of Science and Technology\\
Dept. of Computer and Information Science\\
Trondheim, Norway\\
\textbf{Email}: \{ruocco,heri\}@idi.ntnu.no}

\begin{abstract}
Media sharing applications, such as Flickr and Panoramio, contain a large amount of pictures related to real life events.  For this reason, the development of effective methods to retrieve these pictures is important, but still a challenging task. Recognizing this importance, and to improve the retrieval effectiveness of tag-based event retrieval systems, we propose a new method to extract a set of geographical tag features from raw geo-spatial profiles of user tags.  The main idea is to use these features to select the best expansion terms in a machine learning-based query expansion approach. Specifically, we apply rigorous statistical exploratory analysis of spatial point patterns to extract the geo-spatial features. We use the features both to summarize the spatial characteristics of the spatial distribution of a single term, and to determine the similarity between the spatial profiles of two terms -- i.e., term-to-term spatial similarity. To further improve our approach, we investigate the effect of combining our geo-spatial features with temporal features on choosing the expansion terms. To evaluate our method, we perform several experiments, including well-known feature analyses. Such analyses show how much our proposed geo-spatial features contribute to improve the overall retrieval performance. The results from our experiments demonstrate the effectiveness and viability of our method.
\end{abstract}

\end{frontmatter}

\section{Introduction}
\label{sec:introduction}

The proliferation of web and social media-based photo sharing has not only opened many possibilities but also resulted in new needs and new challenges. Despite recent developments and technological advances within -- e.g., web-based media sharing applications, the continuously increasing amount of available information has made the access to these photos still a demanding task. In general, we can address this challenge by allowing the photo collections to be organized and browsed through the concept of event~\cite{SED2011,RuoccoJournal2012}. Also, most users are generally  familiar with searching photo collections using events as starting points. Thus, aiming at supporting the detection and search of event-related photos, we propose an event retrieval framework to improve the state-of-the-art in real-life event retrieval systems in term of retrieval effectiveness.

Focusing on media sharing applications, an {\em event} refers to "something happening in a {\em specific} place at a {\em specific} time, and tagged with a {\em specific} term"~\cite{RuoccoJournal2012}. With an event-retrieval system, we can assume two types of scenarios: (1) A user directly retrieves media resources related to a particular event; and (2) a user uses a given tagged photo representing an event to retrieve other photos related to any similar events from a large image collection. In this work, we mainly focus on scenario (2). 
Due to their characteristics, pictures in photo sharing applications such as \textit{Flickr}\footnote{See \url{http://www.flickr.com/}} and \textit{Panoramio}\footnote{See \url{http://www.panoramio.com/}} are particularly interesting. For example, most of such pictures are accompanied with contextual metadata and other related information added by users, such as \textit{Title}, \textit{Tags}, \textit{Description}, temporal information represented by the picture capture and upload times, and geolocation. Hence, with photo sharing applications in mind, we study how we can exploit contextual metadata to retrieve event-related pictures. 

The main goals of this work are (1) to build a framework to extract a set of geographical features from geographical raw data of documents or pictures, and (2) to develop an approach to allow effective retrieval of event-based images. Specifically, we develop a set of geographical features that can capture the characteristics of the geographical distributions of social (or user) tags. Further, we investigate how we can combine these features with the state-of-the-art temporal features to improve the retrieval performance of an event-based image retrieval system. Finally, we explore integrating a machine-learning-based approach with our retrieval system. We study how these features can be used in a query expansion framework. Here, we are especially interested in the contributions of the features on the selection of expansion terms from feedback documents.

To this end, we propose a novel framework that improves the retrieval effectiveness of tag-based image search by including the geographical profile of terms. We have developed a new method for extracting spatial features using information about the geographical distribution of tags. Our main idea is to use such features to characterize the clustering tendency of tag terms and the geographical correlation between two geographical distributions of two tags. Spatio-temporal information retrieval is an established field already. However, existing approaches have mainly been concerned with point-of-interests (POI) extraction~\cite{Rae2012} and trajectory mining~\cite{YinCHLH11}. With the constantly increasing number of geotagged pictures -- e.g., in Flickr\footnote{Around 220M of Flickr pictures are geotagged. See also \url{http://www.flickr.com/map/}}, exploring the raw geographical metadata has become increasingly important. 

In summary, the main contributions of this paper are as follows. First, we propose a new robust set of geographical features that can be used (1) to determine the clustering tendency of tags by analysing the geographical structure of their geographical distribution, and (2) to analyse the tag-relatedness between two tags by exploring the correlation between the geographical distributions. To do this, we have developed new measures derived from a well-founded Exploratory Analysis theory from Statistics. More specifically, we adapt the \textit{Ripley's K-function} and \textit{Ripley Cross-K function} ($K$-function and cross-$K$ function for short)~\cite{Ripley1976} as part of our approach to extract the geographical features.  
Second, we show how our features can be incorporated in a machine learning-based query expansion model to improve the ability to select good expansion terms. In addition, we demonstrate how these features can be combined with existing document-based approaches and temporal features to achieve improved retrieval performance. Third, through our experimental evaluation we show the effectiveness and practical feasibility of our approach. This includes comparing with both baseline retrieval models and baseline approaches for geo-temporal tag-relatedness. Fourth, we perform a thorough analysis to show the effectiveness of our proposed geographical features -- in the afore-mentioned machine learning-based query expansion process.  

The rest of this paper is organized as follows. To put our research in a perspective Section~\ref{sec:relatedwork} provides an overview of approaches related to our work. Section~\ref{sec:preliminary} gives an overview of the preliminary theory underlying our approach and defines the problems addressed in this paper.
Section~\ref{sec:proposedframework} presents our proposed geographical features and explains how we extract them. Section~\ref{sec:qe-framework} describes our framework applying these features in a learning-based re-weighting process for a query expansion model. Section~\ref{sec:experimentalsetup} explains our experimental setup. Section~\ref{sec:results} presents the results from our experiments. Finally, in Section~\ref{sec:conclusion} we conclude the paper and outline our future work.

\section{Related Work}
\label{sec:relatedwork}
In the past decades, detection of events from textual document streams and databases has been treated extensively in the literature~\cite{290954,860495}. However, although mining and retrieving pictures related to real-life events is an active field, it is still not a fully mature research domain~\cite{RuoccoJournal2012,Gkalelis2010,Trad2011LSV}. Most related approaches have been aimed at \textit{extracting} events from different types of datasets. To the best of our knowledge, only few works have addressed the problems of \textit{retrieval} of events in connection to media sharing, and many of these approaches were presented in the Social Event Detection (SED) task at MediaEval \footnote{\url{http://www.multimediaeval.org/mediaeval2011/}}~\cite{SED2011}, where the main objective was to propose event retrieval systems for Flickr pictures.

A research area closely related to ours is pseudo-relevance feedback. Generally speaking, pseudo-relevance feedback refers to techniques to average top-retrieved documents to automatically expand an initial query. It has been studied widely in information retrieval both to extend existing retrieval models~\cite{Lavrenko2001,Zhai2001,Tao2006,Cao2008}, and as part of query expansion frameworks~\cite{Carpineto2001,Rocchio2790518}.
Specifically, \citet{Lavrenko2001} and \citet{Zhai2001} propose two methods -- the  \textit{Relevance Model} and the \textit{Mixture Model}, respectively -- to include feedback information in the Kullback-Leibler (KL) divergence retrieval model~\cite{Lafferty2001}. The idea is to estimate a new query model using terms in the top-$k$ retrieved documents, also called pseudo-relevant feedback documents to update an existing query model. Experiments have shown that these approaches are indeed able to improve the standard retrieval models with respect to retrieval effectiveness~\cite{Lv2009}. This has also been the main motivation for including them in our study.
 
\citet{Cao2008} present a classification approach to automatically select \textit{good} expansion terms from a set of candidate terms from the pseudo-relevant documents. To do this, they train a classifier using a set of \textit{good} and \textit{bad} candidate expansion terms represented by feature vectors. Such feature vectors consist of traditional statistical features based on the distribution of the terms both in the whole collection, and the set of (pseudo) relevant documents. \citet{Lin2011} propose an extension of this work by applying a learning-to-rank approach for training and classifying the candidate expansion terms. They show that they can improve the retrieval effectiveness by using social annotation from external tagged resources, such as the {\tt de.li.cio.us}\footnote{\url{http://www.delicious.com/}} social bookmarking web service, as a source for extracting useful expansion terms. The use of social annotation as source for improving the retrieval performance has also previously been investigated by \citet{Zhou2008}. These approaches are related to ours in that we also use classification to select good expansion terms. Their main differences with our approach are that none of them applies either temporal, geo-spatial or geo-spatio-temporal features. 

As discussed later in this paper, we are interested in investigating the contributions of the temporal characteristics of a term in a pseudo relevance feedback context. Within event retrieval, the usefulness of temporal information is evident. Also within general information retrieval, results from existing work have proven its usefulness. For example, \citet{5590248} and \citet{Jones2007} show how the temporal profile of queries can be used to improve existing retrieval models; whereas \citet{Keikha2011} and \citet{Whiting2012} propose new temporal-based approaches to improve pseudo relevance feedback based models. Nevertheless, while existing approaches seem to have focused on the temporal aspects only, to fully support event retrieval, we stress the necessity of the spatial profile of social tags,  as well as the temporal profile. To the best of our knowledge, the combination of both temporal and spatial features of social tags to improve the retrieval effectiveness has still not been sufficiently investigated. Only few methods -- e.g., \cite{Radinsky2011,Zhang2012}, incorporate temporal and spatial correlation measures to compute term-to-term relatedness. Specifically, \citet{Radinsky2011} propose a method to improve the semantic relatedness measure of two terms by capturing the correlation between the temporal profiles of tags and concepts associated with the two terms. \citet{Zhang2012}, on the other hand, analyse the tag relatedness by using different correlation measures, based on spatial and temporal co-occurrence. 
In summary, although these approaches are related ours, the way we extract the spatial profiles of tags and apply them in combination with the temporal profile is different. Also, while these approaches were originally developed for textual documents containing much term redundancy that can normally carry the document semantics, image tags usually consist of few unique terms. This makes it more challenging to derive term-based semantic relatedness for image retrieval in general~\cite{Sun2011}, thus further proving the usefulness of our approaches.

\section{Preliminary}
\label{sec:preliminary}
In this section, we first describe the data our approach is based on and define the problem we address. Thereafter, we give an  overview the statistical method our approach are built on.  

\subsection{Data and Problem Definition}
\label{subsec:problemdanddataefinition}
This work mainly focuses on media sharing applications, where resources are usually tagged with terms -- i.e., tags, that describe the content of the resources. Such resources may also have information specifying their geographical locations, expressed in longitude and latitude values, and are referred to as geotagged resources.

Let $\mathcal{D}=\{P_{1}, \ldots, P_{N}\}$ be a set containing $N=|\mathcal{D}|$ resources. Then, assume that each resource $P_{i}$ can be annotated with a set of tag $T_i$, a temporal timestamp $t_i$ and a geotag $\textbf{g}_i=(latitude, longitude)$, such that  $P_{i} = \{\textbf{g}_i, \tau_i, T_i\}$, $i=1, \ldots, N$.  Without loss of generality, we assume our resources to be a set geotagged pictures downloaded from Flickr, that may or may not contain all of the above information at the same time. 
Further, let $\mathcal{E}=\{E_{1}, \ldots, E_{M}\}$, $M=|\mathcal{E}|$, be a set of picture clusters $E_{i} = \{P_{j_{1}}, \ldots, P_{j_{N_{i}}}\}$, $i=1, \ldots, M$, each of which contains images related to the same event. To make our approach as general as possible, we assume that a query picture has only a set of textual tag terms -- i.e., it does not contain any geotags or temporal timestamps. This means that following our setup above, a query picture related to an event $E_{i_{q}} \in \mathcal{E}$  can be expressed as $P_{j_{q}} = \{T_{j_{q}}\}$ -- i.e., $g_{j_q}$ and $\tau_{j_q}$ are not included. For simplicity, we will use $\Q$ to denote a query picture -- i.e., $T_{j_{q}}=\Q=\{q_1, \ldots, q_n\}$, where $n=|\Q|$ and $q_i, i=1,\ldots, n$ are query tag terms.

The problem addressed in this paper concerns how we can effectively retrieve event-related pictures with a query $\Q$, using only the textual tags. First, we investigate how current state-the-art information retrieval methods perform when applied on our dataset, and let the methods serve as the baseline for our experimental evaluation. Second, we study how a query expansion framework using a set of spatial features summarizing the spatial statistics of the distribution related to a tag, and a set of features defining geographical relatedness between two tags can help us improve the retrieval effectiveness. Third and finally, we compare our method with the baseline methods.

\subsection{Exploring Interaction between Spatial Point Patterns}
\label{subsec:exploringinteractionbetweenspatialpatterns}
As mentioned in Section~\ref{sec:introduction}, our approach is based on geo-spatial features for picture tags. To achieve this, we have to build a {\em spatial profile} for each tag. 

Assume now we have a large dataset $\hat{\mathcal{D}}\subseteq\mathcal{D}$ containing  $L=|\hat{\mathcal{D}}|$ geotagged pictures -- i.e., $\hat{\mathcal{D}}=\{\hat{P_1},...,\hat{P_L}\}$ and $\hat{P_i}=\{\textbf{g}_i, T_i\}, i=1, \ldots, L$. Further, let $\mathcal{V} = \{w_{1},...,w_{W}\}$ be the vocabulary with size $W=|\mathcal{V}|$ of the set of social tags used to annotate $\hat{\mathcal{D}}$. Then, to extract the spatial features from each tag $w_{i} \in \mathcal{W}$, we analyse the spatial characteristics for the tags using statistical {\em exploratory analysis}~\cite{Diggle2003}. 

To be able to use and understand the ideas of exploratory analysis translated into our domain, we need to establish two important concepts our approach is founded on: {\em picture point processes} and {\em tag point pattern}.
First, considering Flickr pictures as our geotagged web resources, we model the spatial distribution of pictures taken in a specific geographical area as {\em picture point processes}, which is formally defined as follows:
\begin{definition}[Picture Point Process]%
	\label{def:picturepointprocesses}%
A Picture Point Process is a point process modelling the spatial distribution of pictures taken in a 2-dimensional study region $\mathcal{R}^{2}$. So, any realization of the random variable, $\mathcal{P}$, modelling the Picture Point Process is called Picture Point Pattern.
\end{definition}
Second, for each term $w_{i} \in \mathcal{V}$, we can assume that we have a set of points 
representing the spatial distribution of the tags in a studied region. With this assumption, we derive a so-called \textit{Tag Point Pattern} from Definition~\ref{def:picturepointprocesses} as:
\begin{definition}[Tag Point Pattern]
\label{def:tagpointpattern}
A Tag Point Pattern $\p_{w_{i}}$ -- or just  $\p_i$ for simplicity -- for a tag term $w_i$ is a subset of a Picture Point Pattern $\mathcal{P}$, and is a set consisting of the geographical positions of pictures annotated with $w_{i}$.
\end{definition}

With these definitions, we can now use statistical exploratory analysis to derive the geo-spatial characteristics of image tags. More specifically, we use a tool called \textit{multivariate Ripley K-function}~\cite{Ripley1976} to get the geo-spatial features from the tags. It is used  to study the interaction between two or more spatial point patterns. To help understand how this is done, below is a brief overview of the \textit{multivariate Ripley's K-function}. 

\subsubsection{Multivariate Ripley's K-Function}
\label{subsubsec:ripley-function} 
The Ripley's \textit{K}-function is mainly a tool for analyzing completely mapped spatial point patterns data in a two-dimensional space~\cite{Ripley1976}. Hence, it can be used to determine the spatial distribution patterns of objects in spaces. 

Let $h$ denote a distance and $\lambda$ be  the intensity of a spatial point pattern, then  Ripley's $K$-function, $K(h)$, is defined as~\cite{Ripley1976}:
\begin{equation}
	 K(h) = \lambda^{-1} E[
	 	\hbox{\#  other points within distance $h$ of an arbitrary point]}
	\label{eq:ripley-k}
\end{equation} 

The multivariate Ripley's $K_{ij}(h)$ function is a generalization of $K(h)$, and is used to analyse the characteristics of an isotropic spatial point process. It contains information about clustering and dispersion of point patterns at different distance scales $h$. The multivariate form aims at answering questions regarding the interaction between two or more point patterns -- i.e., bivariate or multivariate point patterns. It is specified as follows~\cite{Ripley1976}: Let $\lambda_{i}$ and $\lambda_{j}$ be the intensity of the spatial point patterns $\p_i$ and $\p_j$, and assume $\lambda_{i}$ and $\lambda_{j}$ being constant throughout $\mathcal{R}^2$. Then, 
\begin{eqnarray}
	 K_{ij}(h)  & = & \lambda_{j}^{-1} E[
	 	\hbox{\# points of type $i$ within distance $h$} \nonumber \\ 
	 	&& \hbox{from an arbitrary point $j$].}
	\label{eq:multivarkfunctHom}
\end{eqnarray} 
Translated to our application, a point here would be a geographical position of a picture. Restricting to the case of two point patterns, we have four K functions: two \textit{self-K functions} $K_{11}(h)$, $K_{22}(h)$, and two \textit{cross-K functions} $K_{12}(h)$, $K_{21}(h)$. The following is most used estimation of $K_{ij}(h)$, as proposed by \citet{Ripley1976}:
\begin{equation} 
	 \hat{K}_{ij}(h) = \frac{1}{\hat{\lambda}_{i}\hat{\lambda}_{j}A} \sum_{k}\sum_{l} I_{h}(d_{i_{k}j_{l}}),
	 \label{eq:multivarkfunctHomEst}
\end{equation}
where $d_{i_{k}j_{l}}$ is the distance between a \textit{k}-th point of type $i$ and a \textit{l}-th observed point of type $j$. $ I_{h}(d_{i_{k}j_{l}})$ is an indicator, such that  $I_{h}(d_{i_{k}j_{l}})=1$, if $d_{i_{k}j_{l}} \leq h$; and $I_{h}(d_{i_{k}j_{l}})=0$, otherwise. $\hat{\lambda}_{i} = n_{i}/A$ and $\hat{\lambda}_{j} = n_{j}/A$ are the intensity of the two spatial point patterns as the rate between the number of points and the considered area $A$.

The above four $K_{ij}$ functions are used in the exploratory analysis to study the relationship between two spatial point patterns. For example, in the \textit{independence approach} proposed by \citet{Lotwick1982}, the null model assume that two spatial point patterns are generated by two different and independent spatial processes. Under this independence assumption, with the bivariate form or the cross-\textit{K} function, $K_{12}(h) = \pi h^2$. From this, the empirical/estimated cross-\textit{K} function $\hat{K}_{ij}(h)$ calculated on the spatial point patterns, $\p_i$ and $\p_j$, can be compared with the null model to determine the distribution characteristics between the two point patterns as follows: \textit{Attraction}, if $\hat{K}_{ij}(h) > \pi h^2$; \textit{spatial independence}, if $\hat{K}_{ij}(h) = \pi h^2$; and \textit{repulsion}, if $\hat{K}_{ij}(h) < \pi h^2$. 

\subsubsection{Cross-D Function}
\label{subsubsec:d-function} 
As can be derived from the above discussion, Ripley's cross-$K$ functions are useful in characterising the distributions of spatial point patterns. However, the graph of the $\hat{K}_{ij}(h)$ function has normally a parabolic curve, which normally makes it  less straightforward to interpret. As a result, a so-called $L$-function is often used instead. An $L$-function is defined as
\begin{eqnarray}
\label{fun:lcrossfunction}
 L_{ij}(h)=\sqrt{\frac{K_{ij}(h)}{\pi}}.
\end{eqnarray}
 
Using the same assumption of independence of spatial point patterns as before, we get $L_{ij}(h) = h$. As with the $K$-function, the empirical values of $L_{ij}(h)$, $\hat{L}_{ij}(h)$, can be used to characterise tag point patterns $\p_i$ and $\p_j$ as follows: $\hat{L}_{ij}(h) > h$ indicates \textit{attraction} between the point patterns, $\hat{L}_{ij}(h) = h$ shows spatial \textit{independence}, whereas $\hat{L}_{ij}(h) < h$ means \textit{repulsion}. To further facilitate our interpretation,  we normalize the cross-$L$ function again to get a so-called {\em D-function} for two tag point patterns. Based on the empirical cross-$L$ function, the $D$-function is given by 

\begin{equation} 
\hat{D}_{ij}(h) = \hat{L}_{ij}(h) - h.
\label{eq:corrs-D-function}
\end{equation}
 Again, we can use $\hat{D}_{ij}$ to characterize the two tag point patterns $\p_i$ and $\p_j$ as follows: $\hat{D}_{ij}(h)>0$ indicates attraction between $\p_i$ and $\p_j$; $\hat{D}_{ij}(h)=0$ means we have  independence between $\p_i$ and $\p_j$; whereas $\hat{D}_{ij}(h)<0$ implies repulsion between $\p_i$ and $\p_j$. In the rest of the paper, assuming $\p_i \neq \p_j$, we refer this function to as \textit{cross-D function}.

\begin{figure*}[!ht]
	\centering
	\includegraphics[width=5.6cm]{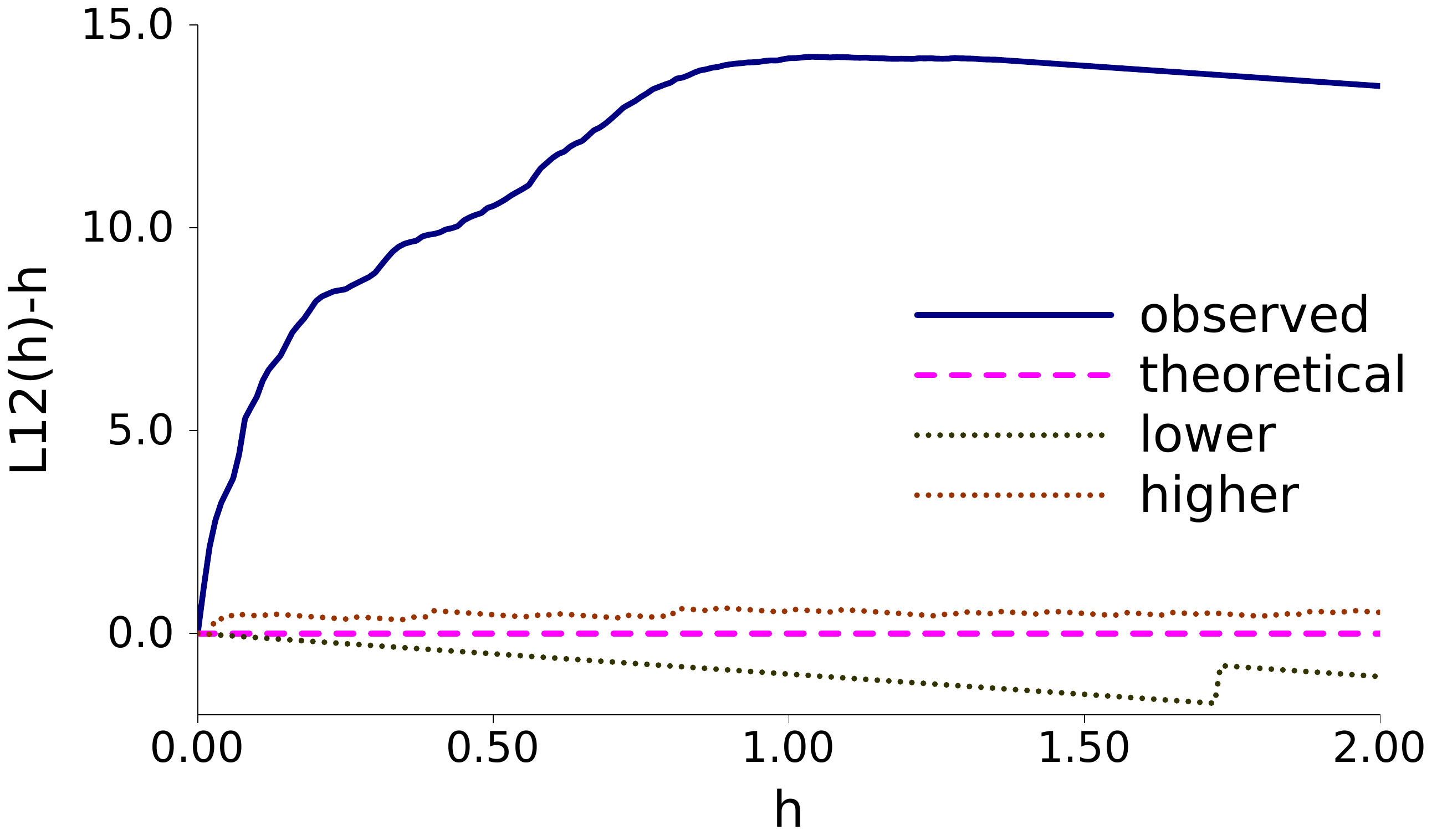}
	\caption{The empirical ({\em observed}) cross-D function $D_{12}(h)$ of the tag point patterns for \texttt{Old Naval College} and the tag \texttt{University of Greenwich} as a function of distance (in km). The  confidence envelopes (95\%) represented by its upper ({\em higher}) and {\em lower} borders, for the {\em theoretical} cross-D function under complete spatial randomness (CSR) are also shown.}
	\label{fig:example-dcross-function-graph}
\end{figure*}

\para{Example [Cross-D function]:}
\begin{figure*}[!hb]
	\centering
	\begin{tabular}{ccc}
				\includegraphics[height=4.8cm]{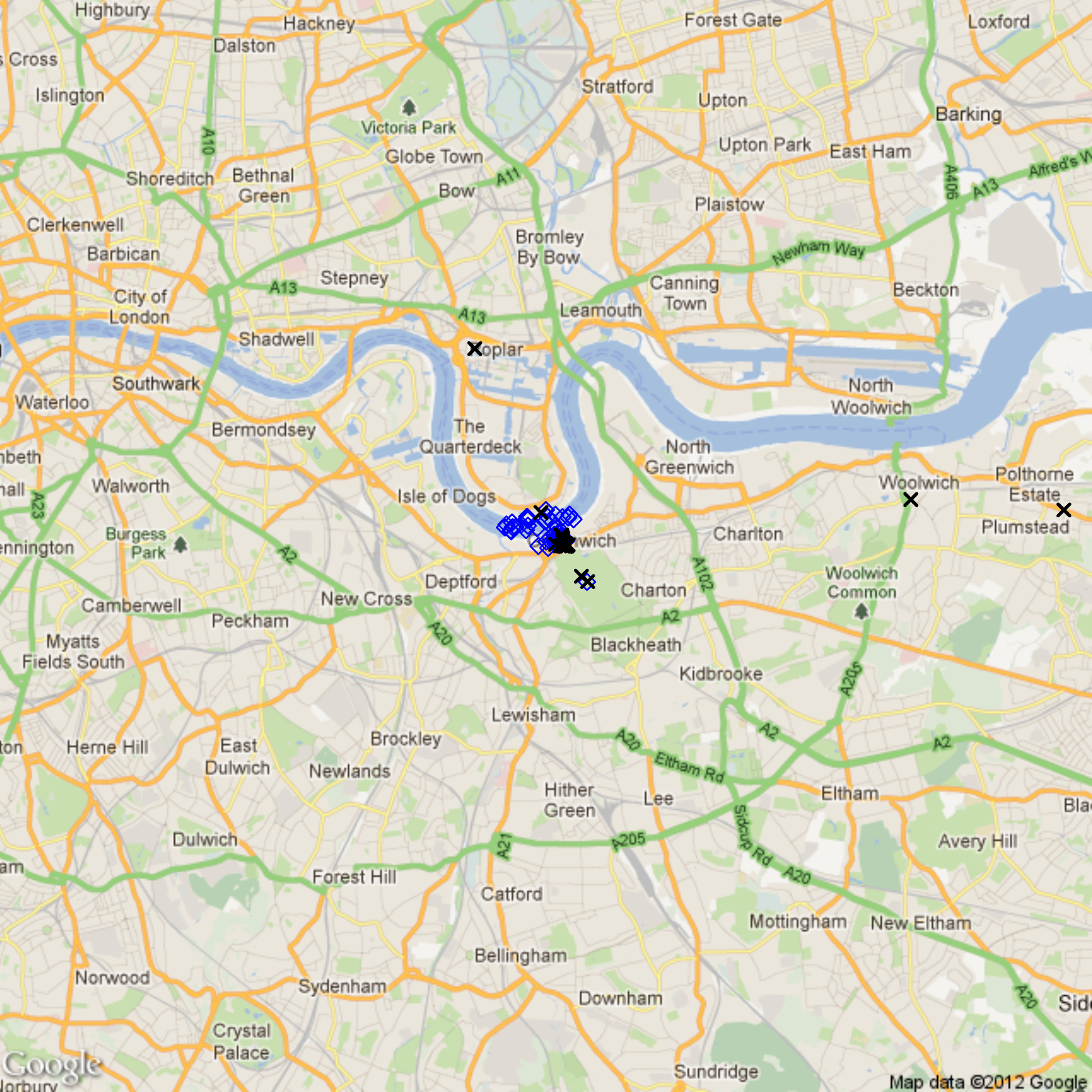}~~
                &
				~~\includegraphics[height=4.8cm]{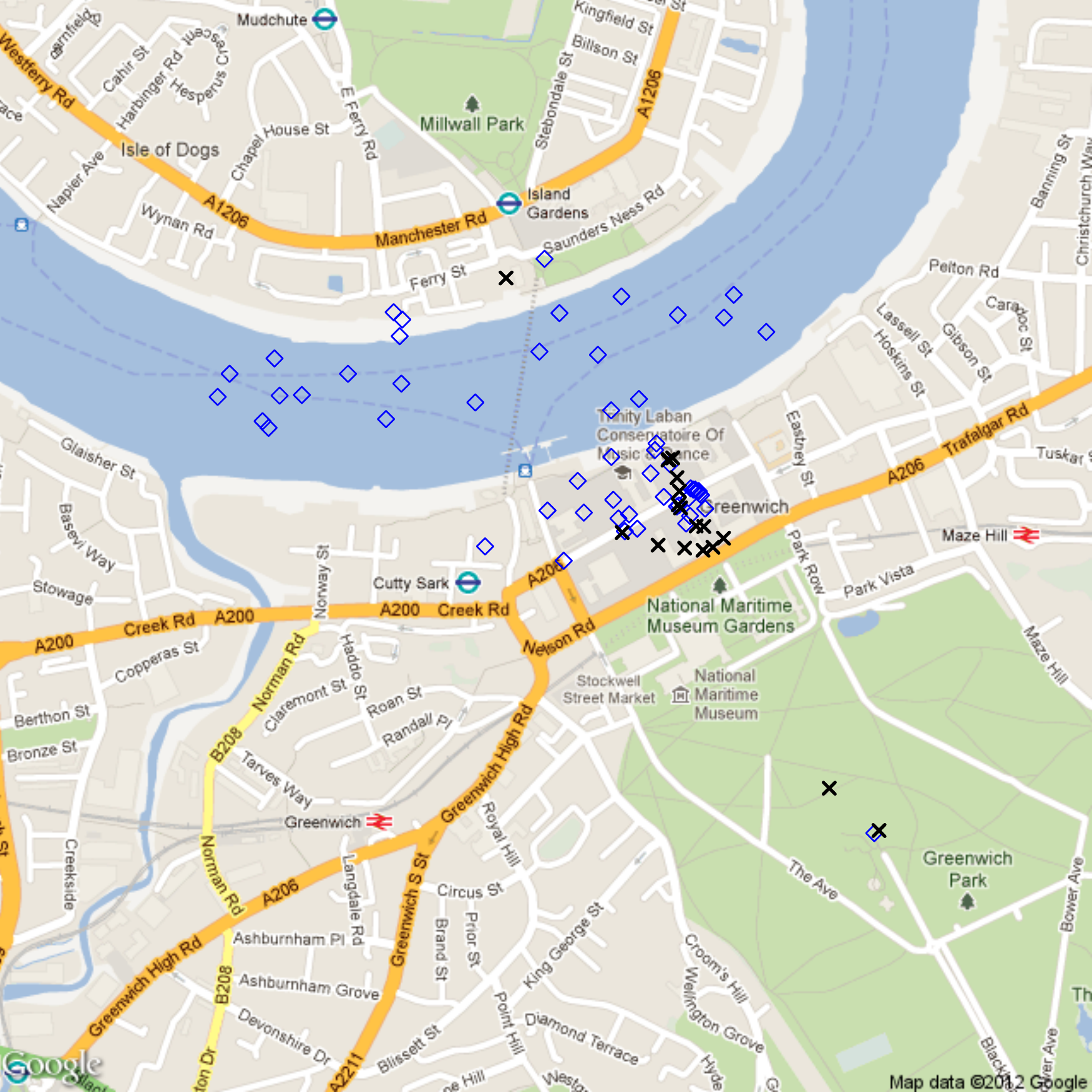}\\ 
   			\centering (a) & (b) 
   			
   	\end{tabular}
	\caption{Spatial distribution of the Tag Point Patterns related to the tag \texttt{Old Naval College} and the tag \texttt{University of Greenwich} at two different zooming (a) and (b)} 
	\label{fig:example-dcross-function}
\end{figure*}
To explain our ideas, assume we have \texttt{"Old Royal Naval College"} and \texttt{"University of Greenwich"} as two specific tags, both referring to areas in London. Then, consider a cross-L function $L_{12}$ between two tag point patterns, $\p_1$ and $\p_2$, as specified in Definition \ref{def:tagpointpattern}, related to these two tags, respectively. A general observation is that the University of Greenwich\footnote{See \url{http://en.wikipedia.org/wiki/University_of_Greenwich}} is located  within the area of the Old Naval College\footnote{See \url{http://en.wikipedia.org/wiki/Old_Royal_Naval_College}}. Thus, although the tags are syntactically different, they are connected and refer to the same geographical entity. Within our spatial statistics, this means that pictures tagged with \texttt{"Old Royal Naval College"} are spatially {\em attracted} to  pictures tagged with University of Greenwich (See Figure~\ref{fig:example-dcross-function}a and \ref{fig:example-dcross-function}b). To further illustrate this relatedness, consider the corresponding cross-D function $D_{12}(h)$ in Figure~\ref{fig:example-dcross-function-graph}, varying the values of $h$ between 0 and 2 km. Using the statistical test described above, we can check the validity of our observation about the spatial attraction among the studied point patterns. As can be seen in Figure~\ref{fig:example-dcross-function-graph},   the graph of $D_{12}(h)$ (denoted as "observed" in the figure) is greater than the upper envelope (denoted as "higher" in the figure), at all values of $h$\footnote{We computed the envelope by simulating the random labelling with a null model and $99$ simulations.}. Hence, based on our distribution "rules" we have {\em "attraction"} between the two point patterns. 

In the following, we elaborate on how we extract our set of features based on the spatial characteristics of a tag point pattern, and the interaction between two spatial point patterns derived from the cross-D function.

\section{Exploring the Spatial Distribution of Tags}
\label{sec:proposedframework}
Recall that the primary goal of this work is to find effective ways to exploit the spatial characteristics of tags to improve the retrieval performance. To achieve this, we investigate applying methods from spatial statistics to explore the spatial distribution of tags. In brief, we apply a collection of features derived from the bivariate Ripley's cross-$L$ function presented in Eq.~\ref{fun:lcrossfunction} and the Ripley's $L$-function for a single tag point pattern. To show how we do this, in Section~\ref{subsec:singletermspatialfeatures} we present our method for extraction of general spatial features of tags, including both single and term-to-term spatial features. In Section~\ref{subsec:norderspatialfeatures}, we focus on special tags, such as tags describing point-of-interests, and introduce a method to extract the spatial features for such tags.  

\subsection{Single and Term-to-Term Spatial Features} 
\label{subsec:singletermspatialfeatures}
We divide the spatial characteristics for tag terms into two main classes: (1) \textit{single-term spatial features},  which determine the aggregation tendency of a single tag spatial point pattern;  and (2) \textit{term-to-term spatial similarity} features, which are related to the geographical similarity between the spatial profiles of two considered tags $w_{i}$ and $w_{j}$. In the following we explain how we extract both these features. 

Assume we have a scale interval $S= [0 \ldots R]$ in kilometres, and that we divide the set of the induced intervals into $K$ discrete and equidistant points ${h_{k}}$,  $k=1, \ldots K$. To extract both the single and term-to-term spatial features, we will use the $D$-function from Section~\ref{subsubsec:ripley-function}, estimated over this interval. 

To capture the clustering tendency of tag point patterns for single terms, in \cite{ruoccoLBSN2012} we introduced two features called $\hat{I}_{SUM}$ and $\hat{I}_{MAX}$ estimators. $\hat{I}_{SUM}$ is computed by extracting the positive area within the intersection between the $D$-function and the curve representing the null hypothesis; whereas $\hat{I}_{MAX}$ is the maximum distance between the $D$-function and the null hypothesis curve.   If we assume that $\hat{D}_{i}$ represents the estimated value of our $D$-function for a tag point pattern $p_i$. Then, for a given tag term $w_i$,
\begin{eqnarray}
 \hat{I}_{SUM}(w_{i}) =& \sum_{k=1}^{K}  \left[ \frac{\hat{D}_{i}(h_{k})}{\sqrt{Var(\hat{D}_{i}(h_{k}))}} \right] ~~\mbox{and}\\ 
 \hat{I}_{MAX}(w_{i}) =& \max\limits_{k=1, \ldots K} \left( \frac{\hat{D}_{i}(h_{k})}{\sqrt{Var(\hat{D}_{i}(h_{k}))}}  \right).
\end{eqnarray}
In other words, $\hat{I}_{SUM}(w_{i})$ is computed by summing the difference between the $D$-function and the null hypothesis. A high value of $\hat{I}_{SUM}(w_{i})$ means that there is a strong {\em aggregation} among pictures that are annotated with $w_i$ and connected to the tag point pattern $p_i$. Further, $\hat{I}_{MAX}(w_{i})$ is calculated by estimating the maximum normalized distance between the $D$-function and the null hypothesis. Hence, it determines the highest positive difference between the $K$-function of the tag point pattern that the $D$-function was derived from and the null hypothesis. A high value of $\hat{I}_{MAX}(w_{i})$ means that the tag point pattern $p_i$ contributes to a high degree of {\em clustering}.

For the bivariate case, we can do similar estimation of the attraction tendency of two tag point patterns as follows: 
\begin{eqnarray}
 \hat{I}_{SUM}(w_{i},w_{j}) =& \sum_{k=1}^{K}  \left[ \frac{\hat{D}_{ij}(h_{k})}{\sqrt{Var(\hat{D}_{ij}(h_{k}))}} \right]~~\mbox{and}\\
 \hat{I}_{MAX}(w_{i},w_{j}) =& \max\limits_{k=1, \ldots K} \left( \frac{\hat{D}_{ij}(h_{k})}{\sqrt{Var(\hat{D}_{ij}(h_{k}))}}  \right),
\end{eqnarray}
where $w_{i}$ and $w_{j}$ are two specific tags with their tag point pattern $p_i$ and $p_j$.

Our initial studies have shown the potentials and the usefulness of the above estimators~\cite{ruoccoLBSN2012}. To apply them in retrieval settings, however, we have to make them more generic, and introduce two new concepts: the \textit{Relative Discrete Positive Area (RDPA)} and \textit{Relative Discrete Maximum Distance (RDMD)}. The main idea is to extend $\hat{I}_{SUM}(w_{i})$, $\hat{I}_{SUM}(w_{i},w_{j})$, $\hat{I}_{MAX}(w_{i})$ and $\hat{I}_{MAX}(w_{i},w_{j})$ by including their behaviour at different scales, and not only at a fixed scale. So, let $\hat{g}_{Sum}$ denote the function representing the relative discrete positive area between the $D$-function and the null hypothesis in a given scale interval, and assume $\hat{g}_{Max}$ represents the maximum distance within the same considered interval. Then, $\hat{g}_{Sum}$ and $\hat{g}_{Max}$ are computed as follows: 
\begin{eqnarray}
 \hat{g}_{Sum}(w_{i},[h_{f},h_{g}]) =& \sum_{k=f}^{g}  \left[ \frac{\hat{D}_{i}(h_{k})}{\sqrt{Var(\hat{D}_{i}(h_{k}))}} \right]~~\mbox{and}\label{eq:rdpa}\\
 \hat{g}_{Max}(w_{i},[h_{f},h_{g}]) =& \max\limits_{k=f, \ldots g} \left( \frac{\hat{D}_{i}(h_{k})}{\sqrt{Var(\hat{D}_{i}(h_{k}))}}  \right),
 \label{eq:rdmd}
\end{eqnarray}
where $f$ and $g$, with $f<g$, are two indexes related to two points $h_{f}$ and $h_{g}$ of the scale interval $S$. Note that if $f=1$ and $g=K$, then $\hat{g}_{Sum}(w_{i},[h_{f},h_{g}]) = \hat{I}_{SUM}(w_{i})$ and $\hat{g}_{Max}(w_{i},[h_{f},h_{g}]) = \hat{I}_{MAX}(w_{i})$. In conclusion, the generalization captures more features, which divide and summarize the spatial characteristics over more sub-intervals within the original scale interval. 

For the bivariate case, we apply a similar approach, and compute $\hat{g}_{Sum}(w_{i},$ $w_{j},$ $[h_{f},h_{g}])$ and $\hat{g}_{Max}(w_{i}, w_{j}, [h_{f},h_{g}])$ by replacing the $D_{i}$ function with $D_{ij}$.

\subsection{N-order Spatial Features} 
\label{subsec:norderspatialfeatures}
The features in Eq.~\ref{eq:rdpa} and Eq.~\ref{eq:rdmd} estimate the deviation of the $D$-function of the  tag point pattern (or the two tag point patterns) from the null hypothesis -- i.e., the spatial randomness for a single tag point pattern, and the spatial independence between two tag point patterns, respectively. In addition to this, in our study we observed that for some tags representing point-of-interests, the curve of the $D$-function related to a tag point pattern tends to be steeper within a short scale sub-interval. Therefore, to also capture such a characteristic, we propose a set of features, called \textit{first order spatial features} that can extract the information on the shape of the curve of the $D$-function.  In Geometry, the \textit{derivative} $f'(x)$ of a source function $f(x)$ can generally be used to determine the slope coefficient of the tangent of the source curve at a point $x$. Using this as a starting point, our idea is to analyse the derivative function of the $D$-function for each sub-interval. Since the $D$-function is discrete over the scale values $h_{k}$, $k=1, \ldots, K$, we apply the discrete equivalent of the derivative function, or more specifically the \textit{forward finite difference}~\cite{smith1985numerical}, as follows:
\begin{eqnarray}
	\hat{D}'_{i}(h) = \Delta_{l,m} \hat{D}_{i}(h) = \hat{D}_{i}(h_{l}) - \hat{D}_{i}(h_{m}), ~~\forall h_{l}<h_{m}, 
	\label{eq:finite-difference}
\end{eqnarray}
where $h_{l}$ and $h_{m}$ are two specific scale points. Note that the value of $\hat{D}'_{i}(h)$ is positive at each scale point where the $D$-function increases, but negative at all scale points where $D$-function decreases. Moreover, the higher the positive value of the $\hat{D}'_{i}(h)$ is, the more the intensity of the function increases. Finally, for the bivariate form of the $D$-function, we can compute the derivative of $\hat{D}_{ij}(h)$ as $\hat{D}'_{ij}(h)$ by extending Eq.~\ref{eq:finite-difference} to take into account both $w_i$ and $w_j$.

Besides determining the slope of the $D$-function, we are also interested in knowing about the {\em concavity} of this function at some point $x$. This gives us more information about the structure or the shape of the function, thus providing us more spatial features. We call such features {\em second order spatial features}, which we get by doing further derivation of the function $\hat{D}'_{i}(h)$. As before, we estimate the resulting $\hat{D}"_{i}(h)$ function by finite differences. 
This means that we can extract the spatial features from $\hat{D}'_{i}(h)$, $\hat{D}'_{ij}(h)$ and $\hat{D}"_{i}(h)$, $\hat{D}"_{ij}(h)$ using the positive area and the maximum distance estimators in Eq.~\ref{eq:rdpa} and Eq.~\ref{eq:rdmd}.

In the next section, we show how the spatial features presented above are useful, especially when used in a query expansion framework for event-based image retrieval.

\section{Query Expansion Framework}
\label{sec:qe-framework}
Query expansion techniques have been one of the most studied approaches within the information retrieval field since the work by \citet{Maron1960}. However, new application areas have made query expansion still needed in order to improve the retrieval effectiveness~\cite{carpineto2012survey}. Nevertheless, reinventing query expansion techniques is not the focus of this work, per se. Rather, we use it as a framework to evaluate the effectiveness of our proposed method on event-related image retrieval. In this section, we specifically elaborate on how we use our proposed spatial features within a query expansion framework. In addition, we explain how spatial features can be combined with temporal features for better retrieval performance.  

\subsection{Overview of the Kullback-Leibler Expansion Model} 
A general query expansion model is a post-processing step in a retrieval system that expand and re-weight an original query with terms from top-$k$ retrieved documents that are assumed to be pseudo-relevant. Such top-$k$ retrieved documents are also called feedback documents.

The Kullback-Leibler divergence-based approach (or just KL-divergence) is a query expansion approach that has been proven  effective focusing on retrieval performance~\cite{Lafferty2001}. The main idea with KL-divergence is to analyse the term distributions, and maximize the divergence between the distribution of the terms from the top-$k$ retrieved documents and the distribution of terms over the entire collection. The terms chosen for the query expansion are those contributing to the highest divergence -- i.e., the terms having the highest so-called KL-scores~\cite{Carpineto2001}. 
To compute the KL-score for a specific term $t$ in the feedback documents, the following equation is used~\cite{Carpineto2001}:
\begin{equation}
KL = P_{Rel}(t)\log\left[ \frac{P_{Rel}(t)}{P_{Coll}(t)}\right],
\label{formula:KLQE}
\end{equation}
where $P_{Rel}(t)$ and $P_{Coll}(t)$ are the probability that $t$ appears in the top-$k$ documents and the collection, respectively. $P_{Rel}(t)$ can be estimated by the normalized term frequency of $t$ in the top-$k$ documents, while $P_{Coll}(t)$ can be computed as the normalized frequency of $t$ in the entire collection. This also means that using Eq.~\ref{formula:KLQE}, terms with low probability in the entire collection and high probability on the retrieved top-k documents have the highest KL-score. 

After the expansion terms have been selected, we can proceed to re-weighting the query terms.  A classical approach for this is the Rocchio's algorithm~\cite{Rocchio2790518} using the Rocchio's Beta equation~\cite{Araujo2008}, given by:
\begin{equation}
	\hat{w}(t_{q}) = \frac{tf_{q_{t_{q}}}}{\max{tf_q}} + \beta \frac{w(t_{q})}{\max{w}},
\end{equation}
where $\hat{w}(t_{q})$ denotes the new weight of a term $t_{q}$ of the query, $w(t_{q})$ is the weight from the expansion model -- i.e., $KL_{Div}(t_{q})$, $\max{w}$ is the maximum weight from the expanded weight model, $\max{tf_q}$ is the maximum term frequency in the query, and $tf_{q_{t_{q}}}$ denotes the frequency of the term in the query.

Since KL divergence is currently one of the state-of-the-art query expansion approaches, it has been the natural baseline approach for our experiments.
 
\subsection{Learning-based Query Expansion Framework}
As can be inferred from our discussion in previous sections, the approach proposed in this paper concerns using geo-spatio temporal features in query expansion frameworks. In particular, we develop a learning-based approach to choose good expansion terms and maximize the retrieval performance. 
\begin{figure*}[!ht]
	\centering
				\includegraphics[width=12cm]{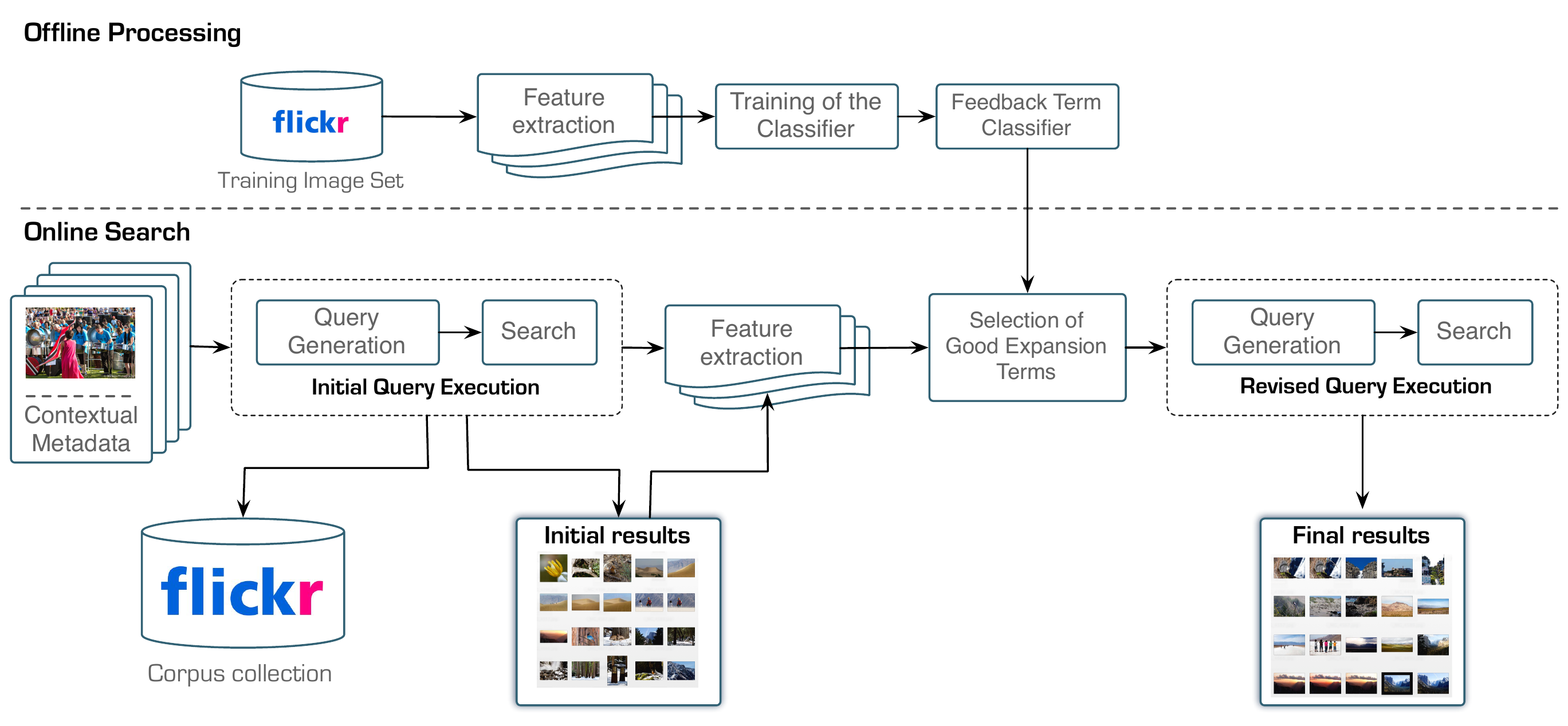}
                \caption{Overview of our supervised learning-based query expansion framework}
                \label{fig:principle}
                
\end{figure*}
%

Figure~\ref{fig:principle} shows the principle behind our approach. As shown in this figure, the process is divided into two main parts consisting of an offline processing and an online search module. In the offline part, we mainly focus on building a classification model for selecting good candidate terms. In the online part, on the other hand, the main focus is on using the model in a search context to select the actual -- previously unseen -- candidate expansion terms.  Algorithm~\ref{alg:retrievalframework} summarizes the steps in the query expansion (QE) process.
\begin{algorithm}[hbtp]
  \caption{Query expansion procedure} 
  \label{alg:retrievalframework}
  \begin{algorithmic}[1]
    \State \textbf{Run query} $\Q$ applying ranking model \textit{r}
    \State \textbf{Get} the set \textbf{D} of top-N relevant docs
    \State \textbf{Extract} unique tags from \textbf{D} and get the candidate expansion term set $\mathcal{E}$
    \For  {$e_{j}$ $\in$ $\mathcal{E}$}
        	\State $\mathbf{\mathcal{X}}$ $\leftarrow$ $ExtractTermFeats(e_{j},\Q)$ \Comment{Sec.~\ref{subsubsec:featureset}}
        	\State $\mathbf{\mathcal{Y}}$ $\leftarrow$ $ExtractTemporalFeats(e_{j},\Q)$ \Comment{Sec.~\ref{subsubsec:featureset}}
        	\State $\mathbf{\mathcal{Z}}$ $\leftarrow$ $ExtractGeoFeats(e_{j},\Q)$ \Comment{Sec.~\ref{subsubsec:featureset}, Sec.~\ref{subsubsec:combiningthespatialfeaturesoveraworlddataset}}
			\State \textbf{Calculate} confidence value $Conf$  \Comment{Sec.~\ref{subsubsec:queryreweightingprocess}, Fig.~\ref{fig:selectiveclassifier}}
			\State \textbf{Combine} \textbf{KL} score and confidence value $Conf$ in a single score $\rightarrow$ $KL_{Final}(\Q,e_{j})$ \Comment{Sec.~\ref{subsubsec:queryreweightingprocess}}
   \EndFor
   \State \textbf{Rank} $e_{j} \in \mathcal{E}$ terms according to  $KL_{Final}(e_{j})$ $\rightarrow$  $\mathcal{E}_{Rank}$
   \State \textbf{Re-build} $\Q$ with the top-k terms from $\mathcal{E}_{Rank}$ $\rightarrow$ $\hat{\Q}$
   \State \textbf{Run Query} $\hat{\Q}$ by using ranking model \textit{r} 
  \end{algorithmic}
\end{algorithm}

An important question is: how do we select the candidate expansion terms? To answer this question, recall $\Q=\{q_{1},...,q_{n}\}$ is query consisting of $n$ terms, and $\mathcal{E}=\{e_{1},...,e_{m}\}$ denote the set of candidate terms for the query expansion process. A \textit{good} candidate expansion term $e_{i}$ is a term that improves the retrieval performance of the original query $\Q$. Building on a similar principle as the approach in \cite{Cao2008}, we find $e_{i}$ by computing the improvements in the average precisions (AP). The idea is as follows. First, for any $e_i$ we compute the average precisions gained from running the original query $\Q$. We call this $AP(\Q)$. Then, we calculate $AP(\Q+e_{i})$, which is average precision for the query we get from expanding $\Q$ with a specific candidate term expansion $e_{i}$. Finally, we find out the improvement in term of average precision from the original query to the expanded on by computing
\begin{equation}
\label{eq:apdiff}
AP_{diff}(\Q,e_{i}) = \frac{AP(\Q+e_{i}) - AP(\Q)}{AP(\Q)}.
\end{equation}
In other words, a candidate expansion term $e_i$ is a good term if $AP_{diff}(\Q,e_{i})$ is positive. Otherwise, it is considered as a bad term. In practice, a threshold $\theta$ is used to control the difference value, such that $AP_{diff}(\Q,e_{i})>\theta$ means we have a good expansion term, whereas $AP_{diff}(\Q,e_{i})<\theta$ means $e_i$ is a bad  term. \citet{Cao2008} suggest $\theta=0.005$ as the default threshold. However, because the application area of \cite{Cao2008} is mainly different from ours, we decided to do an empirical study with different classification algorithms to find the optimal value of $\theta$ (see Section~\ref{subsec:classification-accuracy}). 

To perform the actual term selection, we define the selection task as a binary classification problem. The main idea is to learn a classifier to discriminate the good expansion terms from the bad ones. Thus, we use Eq.~\ref{eq:apdiff} as a basis for the learning process, and to define the positive examples for the classifier. 
As we will discuss in Section~\ref{sec:results}, the main advantage with this approach is its effectiveness.  However, to achieve good results, selection of features is a crucial task. Below, we discuss how we select the feature set that can be used to represent each expansion term $e_i$. Thereafter, we explain how we use a classifier to compute a confidence value as function of $e_i$, as part of the retrieval process. Finally, we present a way to combine this value with the baseline KL-score of the same candidate term to re-rank the set of candidate expansion terms $\mathcal{E}$.

\subsubsection{Selecting the Feature Set}
\label{subsubsec:featureset}
Selecting the right set of features has a direct impact on the accuracy of a classification algorithm.  This is also one of the reasons we emphasize the importance of studying the effects of selection of features in the end (retrieval) results. To learn a classifier, we define a vector of features for each candidate expansion term $e$ from the top-$k$ retrieved items, given a query $\Q=\{q_1, q_2, \ldots, q_n\}$. Table~\ref{tab:all-features} lists the features we study in this work. We group them into three sets of features: \textit{term}, \textit{temporal} and \textit{spatial features}. Since our focus is on event-based retrieval, this calls for features beyond those describing document contents only.
\begin{table}[!ht]
	\small
	\tabcolsep 2pt	
	\center
	\begin{tabular}{p{2.3 cm} p{9.5cm}} 
	\toprule  
		\textbf{Feature} & \textbf{Description}  \\ 
				\midrule
		\multicolumn{2}{l}{\textit{Term Features}}		\\
				\midrule
  		 $DF0(e)$ & Raw document frequency. \\ 
		 $DF1(e)$ & Inverse document frequency: $\log ({N}/{DF0})$. \\ 
		$DF2(e)$ & Inverse document frequency smooth: $\log (1+{N}/{DF0})$.	\\ 
		$DF3(e)$ & Probabilistic inverse document frequency: $\log (({N-DF0})/{DF0})$. \\ 
		 $CoOccSingle(e)$ &  Co-occurrence with single query terms: $\log (\frac{\sum_{i=1}^{n}C(q_{i},e)}{n})$, $n = |\Q|$.  	\\ 
		 $CoOccPair(e)$ &  Co-occurrence with pairs of query terms: $\log (\frac{\sum_{(q_{i},q_{j}) \in \Q}C(q_{i},q_{j},e)}{n})$, $n = |\Q|$. \\
				\midrule
					\multicolumn{2}{l}{\textit{Temporal Features}}		\\
				\midrule
  		 $KURT(e)$,  $KURT(\Q+e)$ & The kurtosis value of the time series for the pictures annotated with an expansion term $e$, and  for the pictures annotated with both an  expansion term $e$ and a query $\Q$, respectively.\\ 
		$AC(e)$,\newline$AC(\Q+e)$ & The autocorrelation value of the time series for the pictures annotated with an expansion term $e$, and  for the pictures annotated with both an  expansion term $e$ and a query $\Q$, respectively.	\\ 
		 $CC(\Q,e)$ & The maximum cross-correlation between the time series for the pictures annotated with an expansion term $e$ and the time series for the pictures annotated with a query $\Q$.	\\
		\midrule
			\multicolumn{2}{l}{\textit{Spatial Features}}		\\
				\midrule
		$\vector{D}_{\textit{Max}}(e)$, $\vector{D}_{\textit{Max}}(e+\Q)$ & The vector of the values from the $\hat{g}_{Max}$ function, related to the $D$-function of the spatial point patterns for pictures annotated with a candidate expansion term $e$,  and with both $e$ and $\Q$, respectively. \\ 
$\vector{D}_{\textit{Max}}(e,\Q)$		 &  The vector of the values  from the $\hat{g}_{Max}$ function, related to the cross-$D$-function between tag point patterns associated to $e$ and the tag point patterns of $\Q$.\\ 
		 $\vector{D}_{\textit{Sum}}(e)$, $\vector{D}_{\textit{Sum}}(e+\Q)$  & The vector of the values from the $\hat{g}_{Sum}$ function, related to the $D$-function of  the spatial point patterns  for the pictures annotated with a candidate expansion term $e$, and with both the terms $e$ and $\Q$, respectively. \\
		$\vector{D}_{\textit{Sum}}(e,\Q)$ &  The vector of the values  from the $\hat{g}_{Sum}$ function, related to the cross-$D$-function between tag point patterns associated to $e$ and the tag point patterns of $\Q$.	\\ 
		 \bottomrule
\end{tabular}
\caption{A Summary of the Set of Features} 
\label{tab:all-features}
\end{table}

\para{Term Features ($\mathcal{X}$):} The term features consist of features that are used to characterize a document content. They are chosen based on the hypothesis that terms that contribute to improve the retrieval effectiveness are those being most frequent and distinctive~\cite{Cao2008}. Existing studies suggest using features related to the distribution of the candidate term $e$ in the feedback documents and the whole collection, and those capturing the co-occurrence of $e$ with the terms in the original query $\Q$~\cite{Cao2008,Lin2011}. It is, however, worth noting that these features has mainly been applied in full-text document retrieval, where term redundancy is normal, and thus term frequency would be an important feature. Since a tag generally appears only once for each picture, term frequency as a feature has generally no impact on the classification accuracy. For this reason, in our experiments, our set of term features does not include term frequency but other traditional statistical features such as document frequency (DF)~\cite{BaezaYates11}. As part of evaluating our approach, we will use the term features in implementing the baseline approach for our experiments. This will also allow us to assess how well the features suggested in this work improve the retrieval performance. 

\para{Temporal Features ($\mathcal{Y}$):} Once again, since our focus is on event retrieval, we are interested in capturing how each term in image tags contributes to characterising the images over time periods. Therefore, we need a set of statistical features that represent the temporal distribution of the term in the whole collection. Here, we propose single term features and term-to-term features related to the temporal correlation of the candidate expansion term and the query terms. More specifically, to capture the characteristics of the temporal distribution of a single term, we adopt the concept of \textit{kurtosis} defined as ${\mu_{4}}/{\mu^{2}_{2}}$, where $\mu$ is the mean and $\mu_{j}$ is the $j$-th central moment. Kurtosis were originally proposed by \citet{Jones2007} to capture the dynamics of a time series. It can be used to quantify the probability distribution concentrated in peaks of a time series -- i.e., the "peakedness". In this work, we propose to measure the peakedness for both a single candidate expansion term $e$ ($KURT1$), and the combination of a candidate expansion term $e$ with a term $q_i$ from the original query ($KURT12$).

In addition to this, we are interested in knowing about the randomness of terms over time. A way to detect such a randomness is to use {\em autocorrelation}~\cite{box2013time}. In general, autocorrelation is computed by finding the statistical correlation between two values of the same variable at a given time $t_l$ and another time $t_{l+m}$. Such values can, for example, be the number of occurrences of a term $e$ at specific times. The hypothesis is that bursty events in a time series normally contribute to a high autocorrelation value~\cite{Jones2007}. To capture this, we compute the first order \textit{autocorrelation} of a time series for both a single candidate expansion term $e$ ($AC1$), and the combination of a candidate expansion term $e$ with a term $q_i$ from the original query ($AC12$). 
Finally, to measure the temporal similarity between the time series of two different terms $q_i$ and $e$, we can apply the {\em cross-correlation} measure ($CC$)~\cite{Chien2005,Radinsky2011}. Cross-correlation is computed by assessing the correlation of the frequency of $q_i$ and $e$ to measure the relationship between $q_i$ and $e$. To compute the temporal features, we varied the time windows or bins from one day to seven days, with which seven days gave the best results.
To summarize, we investigate how combining previously proposed temporal features would affect the retrieval performance. These have proven successful in other more general information retrieval approaches, but the way we analyse the effects of their combination within event-related image retrieval haven't been done before. 

\para{Spatial Features ($\mathcal{Z}$):} As explained in Section~\ref{sec:introduction}, the concept of event is strongly related to the spatial dimension -- i.e., {\em geographical location}. We hypothesize that a good expansion term  is {\em spatially correlated} with at least one of the query terms. This is the main reason we study the impact of clustering tendency, with respect to the spatial distribution for the pictures annotated with the candidate expansion terms. As part of this, we compute the spatial features as presented in Section~\ref{subsec:singletermspatialfeatures}. For each pair of terms $e$ and $q_{i}$, we first extract the set of geographical world tiles $\tile_{q_{i},e}$ containing spatial points related to documents annotated with $q_{i}$, spatial points associated to documents annotated with $e$, and those related to documents annotated with both $q_{i}$ and $e$. 
Next, we extract a set of six spatial feature vectors from each tile $\tile_{q_{i},e}$. The first three feature vectors are the vectors computed using $\hat{g}_{Max}$ -- i.e., the \textit{relative discrete maximum distance} function for the specified tag point pattern (see Eq.~\ref{eq:rdmd}), consisting of $\vector{D}_{\textit{Max}}(e)$, $\vector{D}_{\textit{Max}}(e+\Q)$, $\vector{D}_{\textit{Max}}(e,\Q)$. The second set of feature vectors are based on $\hat{g}_{Sum}$ -- i.e., the \textit{relative discrete positive area} function  (see Eq.~\ref{eq:rdpa}), consisting of $\vector{D}_{\textit{Sum}}(e)$, $\vector{D}_{\textit{Sum}}(e+\Q)$, $\vector{D}_{\textit{Sum}}(e,\Q)$. For all the extracted features, we compute the values of the functions by varying the distance values from $0$ to $1~km$, with a step of $0.1~km$. 
Finally, for both the resulting first and second derivative of $\hat{g}_{Max}$ and $\hat{g}_{Sum}$, we perform similar operations as described in Section~\ref{subsec:norderspatialfeatures}. Note that as can be inferred from this, the input query $\Q$ used to extract the features may have varying dimensions. However, this does not cause problem but may only affect the number of spatial points used to build the feature vectors, which is, according to Eq.~\ref{eq:multivarkfunctHom} -- \ref{eq:corrs-D-function}, implicitly decided by the value of the distance scale $h$.

In this work, we study the impacts of  with these features combining the temporal features. In Section~\ref{sec:results}, we analyse their usefulness and importance with respect to improving the retrieval performance. 

\subsubsection{Combining the Spatial Features using the World Dataset}
\label{subsubsec:combiningthespatialfeaturesoveraworlddataset}
Our dataset has been built from a collection of Flickr pictures covering the whole world map. For this reason, the spatial distribution of the pictures is not uniform. To cope with this, we divide the entire world map into a number of tiles. More specifically, we divide the world map into grids with size of one latitude degree and one longitude degree. We span the latitude in the range of $[-180,...,+180]$ degrees, while the longitude in $[-70,...,+70]$, instead of $[-90,...,+90]$ degrees to avoid the Arctic and Antarctic areas, since these areas have normally poor photographic activity. The width of each tile for each (or one) degree of latitude is constant, and has a size of $111$ km, while following the latitude values, the tile heights vary from around $0$ at the poles to around $111$ km at the equator. This would give us in total 50,400 tiles. 
However, to restrict the computation cost, we only consider tiles containing a significant number of pictures -- i.e., more than 1,000 pictures. Let such tiles be significant tiles, denoted by $\tile$.

With this in mind, we extract the spatial feature vectors for a pair of terms $w_{i}$ and $w_{j}$ -- e.g., a query term $q_i$ and an expansion term $e$, as follows. First, let $\tile_i = \{\tile_{i_{1}}, \tile_{i_{2}} \ldots, \tile_{i_{N}}\}$ be the set of $N$ tiles containing pictures tagged with $w_{i}$, and  $\tile_j = \{\tile_{j_{1}}, \tile_{j_{2}} \ldots, \tile_{j_{M}}\}$ be the set of $M$ tiles containing pictures tagged with $w_{j}$. Then, to find a significant tile, $\tile_{ij} \in \tile$, containing pictures tagged with both $w_{i}$ and $w_{j}$, we merge the two sets $\tile_i$ and $\tile_j$.
Finally, to get the feature vectors, for each tile, $\tile_{ij}$, we compute the bivariate $D$-function and the corresponding estimators for the tag point patterns for both $w_{i}$ and $w_{j}$ (see Section~\ref{sec:proposedframework}).

To have a data structure allowing efficient feature extraction operations, we index each tile $\tile_{l}$ as a document composed by the set $\mathcal{W}_{\tile_{l}} =  \{w_{l_{1}}, \ldots, w_{l_{|\tile_{l}|}}\}$ of all tags annotating the pictures from each tile $\tile_{l}$. To do this, we create an inverted index, $\mathcal{I}$, for each tag $w_{l_i} \in \mathcal{W}_{\tile_{l}}$, which we can formulate formally as follows:
\begin{equation}
	\mathcal{I} : \{w_{i} \longrightarrow \{ <\tile_{i_{1}}, tf_{\tile_{i_{1}}}(w_{i})>, <\tile_{i_{2}}, tf_{\tile_{i_{2}}}(w_{i})>,...\} \}_{i}
\end{equation}  
This means that each tag is linked to an inverted list containing the id of the tile and the term frequency, $tf_{\tile_{i_{1}}}(w_{i})$, of a tag, $w_{i}$.

\para{Selection of the Tiles for Spatial Features Extraction}.  To select the tiles for spatial feature extraction, we are mainly interested in the tiles containing pictures that are annotated with both at least one term in $\Q$ and a candidate expansion term $e$. However,  to make the spatial features suitable for our classifier, we select only one tile that is most representative to a specific input query $\Q$. We call this the best tile. To do this, we first run $\Q$ on our dataset. Then, we select the first $K$ geotagged pictures from the resulting ranked list. Finally, we select the tile containing the highest TF-IDF-based ranking score. For simplicity, by treating  tiles as documents, we index and search them using \textit{Solr}\footnote{\url{http://lucene.apache.org/solr/}} search platform. Thus, the resulting list of tiles is ranked using Lucene scores\footnote{\url{http://lucene.apache.org/core/3_6_2/scoring.html}}.

\subsubsection{Query Re-weighting Process}
\label{subsubsec:queryreweightingprocess}
We now explain how we perform the re-weighting process using the sets of features presented in the previous sections.

\begin{figure}[!ht]
	\centering
				\includegraphics[width=12cm]{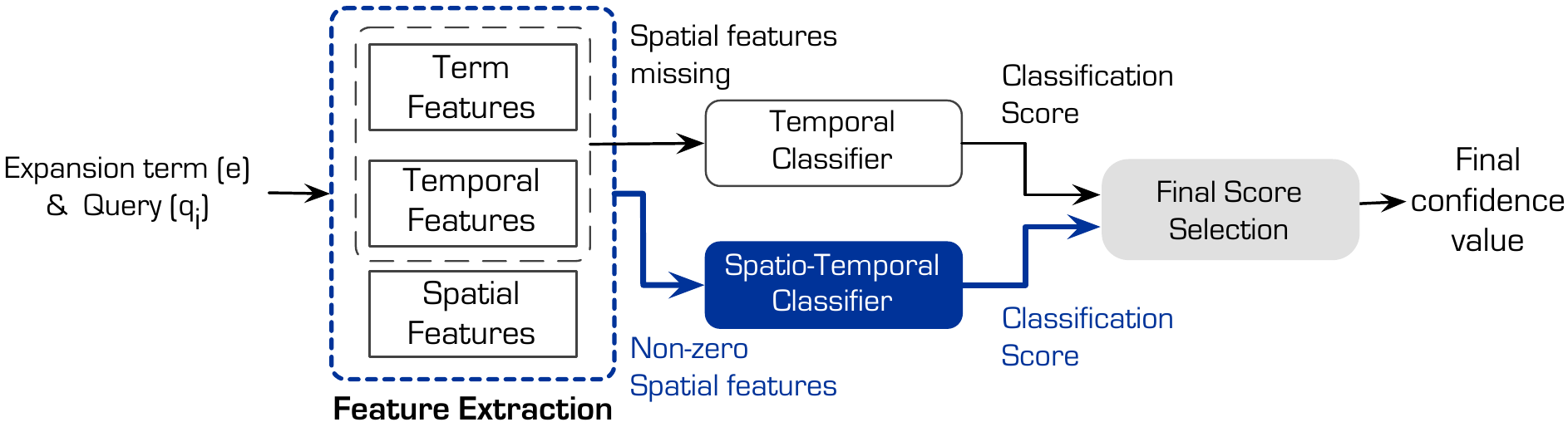}
                \caption{Good expansion term selection process through classification}
                \label{fig:selectiveclassifier}
\end{figure}
Figure~\ref{fig:selectiveclassifier} shows a part of the term selection and re-weighting process. As depicted in the figure, the \textit{Temporal Classifier} is trained with positive and negative examples using only term and temporal features, while the \textit{Spatio-Temporal Classifier} is trained with instances using the complete set of features. Thus, given a query term $q_{i}$ and the candidate expansion term $e$, we first extract the complete set of features. Thereafter, the input instances are classified with both of the classifiers. Finally, a \textit{Final Score Selection} module designates the final confidence value. By default, our system produces scores based on all three feature sets. However, we might have a situation in which we do not have geo-tagged pictures that are annotated with both $e$ and any $q_{i} \in \Q$. Thus, producing the spatial feature vectors from the functions $g_{Max}(\Q+e)$ and $g_{Sum}(\Q+e)$ would be hard. If this happens, then the final score from the  Final Score Selection module is based on the term and temporal features only. 

We call the final confidence score for good candidate expansion terms from our expansion term selection process $Conf(+|e)$. To produce the final Kullback-Leibler (KL) score for the query expansion process, we combine $Conf(+|e)$ with the term-based KL-score as follows:
\begin{equation}
\label{eq:kl-final}
KL_{Final}(e) = \alpha KL(e) + (1-\alpha) Conf(+|e).
\end{equation}
Note that to allow this combination, both $Conf(+|e)$ and $KL(e)$ values are normalized. Here, $\alpha$ is a constant used to decide which component should have the highest contribution. That is, $\alpha=1$ means that we only apply the regular KL-divergence, whereas $\alpha=0$ means we rely entirely on the classification modules to choose good expansion terms. Since we are interested in the impacts of our expansion terms to the retrieval performance, we let both components to have equal contributions to the final score -- i.e., we use $\alpha=0.5$. 

The confidence value $Conf(+|e)$ is computed based on the idea that both the temporal and the spatio-temporal classifiers give their contributions exploring terms over different dimensions, and that they complement, rather than extend each other. With this in mind, $Conf(+|e)$ can be computed as follows:
\begin{equation}
Conf=
\left\lbrace %
\begin{aligned}
    &\mbox{ {\small 0,} }&\mbox{ {\small if $Conf_{T} <0.5$   and  \small $Conf_{ST}<0.5$}}\\
    &\mbox{ {\small $Conf_{T},$} }&\mbox{ {\small  if $Conf_{T}>0.5$   and  \small $Conf_{ST}<0.5$}}\\
    &\mbox{ {\small $Conf_{ST},$} }&\mbox{ {\small  if $Conf_{T}<0.5$   and  \small $Conf_{ST}>0.5$}}\\
    &\mbox{ {\small $\frac{Conf_{T}+Conf_{ST}}{2},$} }&\mbox{ {\small  if $Conf_{T}>0.5$   and  $Conf_{ST}>0.5$}}
\end{aligned} %
\right.
\end{equation}
Here, $Conf_{T}(+|e)$ and $Conf_{ST}(+|e)$ are the confidence values from the Temporal Classifier and Spatio-Temporal Classifier, respectively. The choice of the value $0.5$ as threshold is made based on the fact that $0 \leq Conf(+|e)  \leq  1$ and that we aim  at having final confidence values higher than half the highest possible value. Our experimental results have shown that this is a sensible choice.

\section{Experimental Setup}
\label{sec:experimentalsetup}
In this section we present our dataset and the methodology for our experimental evaluation. 
\subsection{Dataset}
\label{subsec:dataset}
To perform our experiments with tag-based search of event retrieval pictures and to check the feasibility of our approach, we use a large dataset of pictures gathered from Flickr\footnote{We used Flickr API, \url{http://www.flickr.com/services/api/}} covering a time period from 01.01.2006 to 31.12.2010 and without spatial restrictions. This results in a final dataset consisting of 88,257,485 pictures, of which 18,861,585 pictures are without any tags and around $23.5 \%$ are with $1$ to $3$ tags. For relevance judgement we apply the well-established \textit{Upcoming dataset}~\cite{BeckerNG10} as our ground truth. It has also been used previously in other related approaches~\cite{Wang2012SED}. 
Specifically, the Upcoming dataset consists of 270,425 pictures from Flickr, taken between 01.01.2006 and 31.12.2008, each of which belongs to a specific event from the Upcoming event database\footnote{See \url{http://www.cs.columbia.edu/~hila/wsdm-data.html}}. The unique number of events are 9,515. Each event is composed by a variable number of images, varying from $1$ to 2,398 pictures. This large number and the heterogeneity of the included events are the main advantage of the Upcoming dataset, and the main reason we decided to use it. For generality, we merged the Upcoming dataset with the set of other Flickr pictures.

To perform our experiments, we indexed all image tags using Terrier\footnote{See \url{http://www.terrier.org/}}. As part of the dataset preparation, we perform a preprocessing step consisting of tokenization based on whitespace and punctuation marks; stemming, by using the Porter stemmer algorithm~\cite{Porter1980}; and English stopword removal. 

\subsection{Evaluation Methodology}
\label{subsec:evaluation-methodology}
In this section, we briefly explain how we evaluate our approach. First, we present our input query set. Second, we discuss the methods we used as baseline for our experiments. Third, we elaborate on the evaluation metrics we applied.

\subsubsection{Input Query Set} We randomly selected set of $150$ pictures, one for each event cluster in the Upcoming dataset, and use the tags annotating the pictures as queries. We divide this set of queries into two subsets, one subset consisting of $100$ queries that we use to train and evaluate the performance of the classifiers, and another subset consisting of the remaining $50$ queries that are used as the test set to evaluate the retrieval effectiveness of the proposed retrieval framework. For completeness, in Table \ref{tab:inputqueryexample}, we show some example of input queries used in our experiments.

\begin{table}[!ht]
    \vspace*{-0.5\baselineskip} 
    \tabcolsep 2pt
    \small
    \center
    \begin{tabular}{p{3.4cm}p{5cm}}
    \toprule
   \textbf{Query} & \textbf{Event Description}\\
    \midrule
   \multicolumn{2}{l}{\includegraphics[height=7cm]{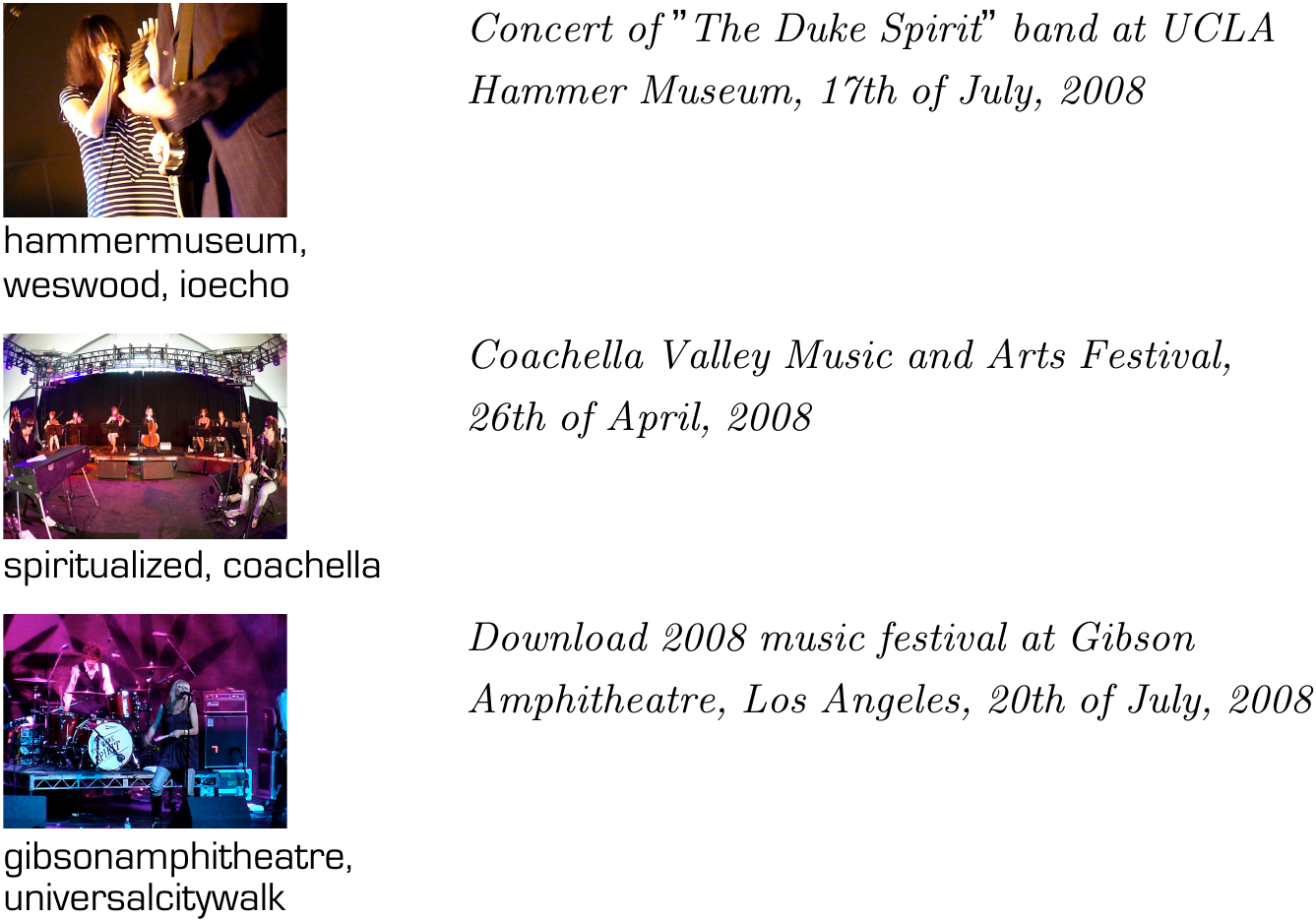}}\\

\end{tabular}
\caption{Example of queries extracted from the Upcoming dataset.}
\label{tab:inputqueryexample}
\end{table}

\subsubsection{Baseline Methods} 
\label{subsec:base-line-methods}
To assess the effectiveness of the retrieval framework, we compare our models with several baseline methods. First, we perform the searching process by using classical retrieval models, including the \textit{Vector Space Model (VSM)}, \textit{Okapi BM25 (BM25})~\cite{Robertson1994}, and the \textit{Language Model (LM)} for information retrieval -- with Jelineck and Dirichlet smoothing. Since BM25 gave the best results in term of effectiveness, we only show the results related to this model. We use the default parameter values $k_{1}=1.2$, $k_{3}=8$ and $b=0.75$ as baseline for our evaluation. As a query expansion model, we use the basic KL-divergence model (\textit{KL}) and the machine learning  approach with the baseline features as proposed \citet{Lin2011} as baseline (\textit{KLML}). For simplicity and readability, we only show the results of \textit{KL} since we observed that the MAP values of \textit{KLML} are comparable with the MAP values of \textit{KL}. We compare the baseline approaches with our proposed methods, first by comparing them with a query expansion framework applying a classifier trained with the combination of {\em terms} and {{\em temporal features} (\textit{KL\_T}); and then a framework with a classifier learned with the combination of {\em terms}, {{\em temporal}, and {\em spatial features} (\textit{KL\_ST}).
Note that in addition to the above models, we also experimented with the Mixture Model~\cite{Zhai2001} and the Relevance Model~\cite{Lavrenko2001}, also incorporating the feedback documents in the ranking score computation. However, the results from these experiments were, though comparable, worse than those from the \textit{BM25+KL} query expansion models.  Thus, for simplicity we did not include the results from these experiments.

\subsubsection{Comparison with Related Work}
\label{subsec:comparison-related-work}
To have a fair comparison with similar approaches, we implemented the geo-temporal tag relatedness by \citet{Zhang2012}, which we, from now on, refer to as  {\em ZKYC\cite{Zhang2012}} for simplicity. As with our approach,  with \ZKYC~the similarity between two tags is computed by comparing their temporal and geographical distributions with so-called \textit{geo-spatial}, \textit{temporal}, and \textit{geo-temporal} similarity measures. First, they quantize the world map (space) into $m$ tiles of $1$ degree, and the time into $n$ temporal bins of two weeks. Then, they extract the tag features based on the three measures using vectors of numbers of users applying a tag in each bin. This means that for a specific tag the geo-spatial feature vector  contains $m$ elements of numbers of users applying that tag in each bin;  the temporal feature vector contains $n$ elements of numbers of users applying the tag in each bin; and the geo-temporal feature vector or matrix  contains $m \times n$ elements of the counts of unique users tagging a picture within the geo-temporal bin. All vectors are normalized with $l^2$-norm. \citet{Zhang2012} get the similarities between two tags by computing the euclidean distance between the two corresponding feature vectors. 

As can be inferred from this, the main difference between our approach and  ZKYC\cite{Zhang2012} is the geographical and geo-temporal features used and how they are extracted. Specifically, with ZKYC\cite{Zhang2012}, the geographical feature vector related to a tag is static, and representing the distribution of the tag over a single size of bin; whereas basing our approach on the Ripley $K$-function enable characterizing the geographical distribution of tags over non-fixed geographical scales. As discussed previously, the Ripley $K$-function also allow us to extract the clustering properties of tags. In our experiments, we pay special attention to how this difference affects the retrieval performance. Specifically, we consider a range between 0 km and 3 km of scales when computing the $K$-function. According to \citet{Zhang2012}, the geo-temporal features yielded the best results. For this reason, we only compare our approach using with the one applying the geo-temporal features.  To incorporate this relatedness in a retrieval framework and compare it with our approach, we define a ranking equation equivalent to Eq.~\ref{eq:kl-final} as $\alpha KL + (1-\alpha) rel_{geo-temp}$, where $\alpha$ is a  constant deciding the contribution of the components, $KL$ is the KL-score and $rel_{geo-temp}$ denotes the geo-temporal tag relatedness score. We tune and select the best value of the parameter $\alpha$ over a set of 50 queries.

\subsubsection{Evaluation Metrics} To evaluate the retrieval performance of all the models, we use Mean Average Precision (\textit{MAP}), a widely used evaluation metric within information retrieval~\cite{BaezaYates11}. We compute our MAP values based on 1,000 retrieved documents (images). To make sure that any improvements are statistically significant, we perform paired two-sample one-tailed t-tests at $p<0.05$ or 95 \% confidence interval. Any stated improvements in this paper are all statistically significant, unless otherwise specified.
\section{Results}
\label{sec:results}
In this section we perform two different analyses. First, we study the impact of using our temporal and spatial features on training classification algorithms. Second, we investigate the effectiveness of using the temporal and spatial features in a classifier with an optimal feature selection procedure.

\subsection{Classification Accuracy}
\label{subsec:classification-accuracy}
As part of the process of designing a good classifier for selection of good expansion terms, we investigated which classifier is suitable for our application. We evaluated several existing classifiers with respect to their classification accuracy, and selected the classifier yielding the best accuracy. Specifically, we tested our method using \textit{Na{\"i}ve Bayes} classifier, \textit{Support Vector Machine} (\textit{SVM}), \textit{C4.5} decision tree (also named \textit{J48}) and \textit{Random Forest}. We used Weka~\cite{Hall2009} machine learning toolkit with default parameter settings to test the classifiers. The training set was composed by a set of 1,000 terms, equally divided into good and bad terms. These were obtained by randomly selecting the feedback terms from the results of running the queries using the training set. 

To perform a thorough evaluation, we calculated the \textit{accuracy}, \textit{precision} and \textit{recall} values for each classifier, with a leave-one-out cross validation. We performed the test for five different training sets that we obtained by selecting a positive and a negative class using different values of threshold $\theta$ (see also Section \ref{sec:qe-framework}). The $\theta$ values we selected were  0.001, 0.005, 0.01, 0.05, 0.1, and 0.5. 

We summarize the averaged results in Table~\ref{tab:classificationaccuracy}. Here, "$+$" is the positive class containing the good candidate terms, whereas "$-$" denotes the negative class holding terms considered bad expansion terms. From these results, we can observe that the general performances of the classifiers are good using the proposed set of features. The overall best result was gained by using Random Forest classifier, with an accuracy of around $95\%$. The precision for the classification of the good terms was $93\%$ and the recall was as high as $97.56\%$.
\begin{table}[!ht]
\tabcolsep 2pt
\small
\center
\begin{tabular}{lccccc}
\toprule  
\multicolumn{2}{c}{\textbf{ }} & \multicolumn{2}{c}{\textbf{Precision}} & \multicolumn{2}{c}{\textbf{Recall}} \\
\midrule

&   \textbf{Accuracy} (\%) &  $+$ & $-$ &  $+$ & $-$ \\ 
\midrule \textbf{Na{\"i}ve Bayes} & 59.12 & 0.6102 & 0.6092 & 0.6180 & 0.5644 \\ 
\midrule  \textbf{SVM} & 69.22 & 0.6802 & 0.7072 & 0.7288 & 0.6556 \\ 
\midrule  \textbf{C4.5/J48} & 91.52 & 0.8870 & 0.9490 & 0.9536 & 0.8768 \\ 
\midrule  \textbf{Random Forest} & \textbf{94.98} & \textbf{0.9288} & \textbf{0.9736} &\textbf{0.9756} & \textbf{0.9240} \\ 
\bottomrule
\end{tabular}
\caption{Comparison of the classification performances. The best scores in each column are type-set boldface.}
\label{tab:classificationaccuracy}
\end{table} 

We now analyse the behaviour of the accuracy value of the four proposed classifier over the different $\theta$ values. The results is summarized in Figure \ref{fig:accuracylambda}. Here, we can observe that the higher the threshold value is, the more the accuracy of the classifier increases. Moreover, both J48 and Random Forest (RF)  outperformed the Na{\"i}ve Bayes (NB) and SVM, with high margin. 
\begin{figure}[!ht]
	\centering
				\includegraphics[height=4cm]{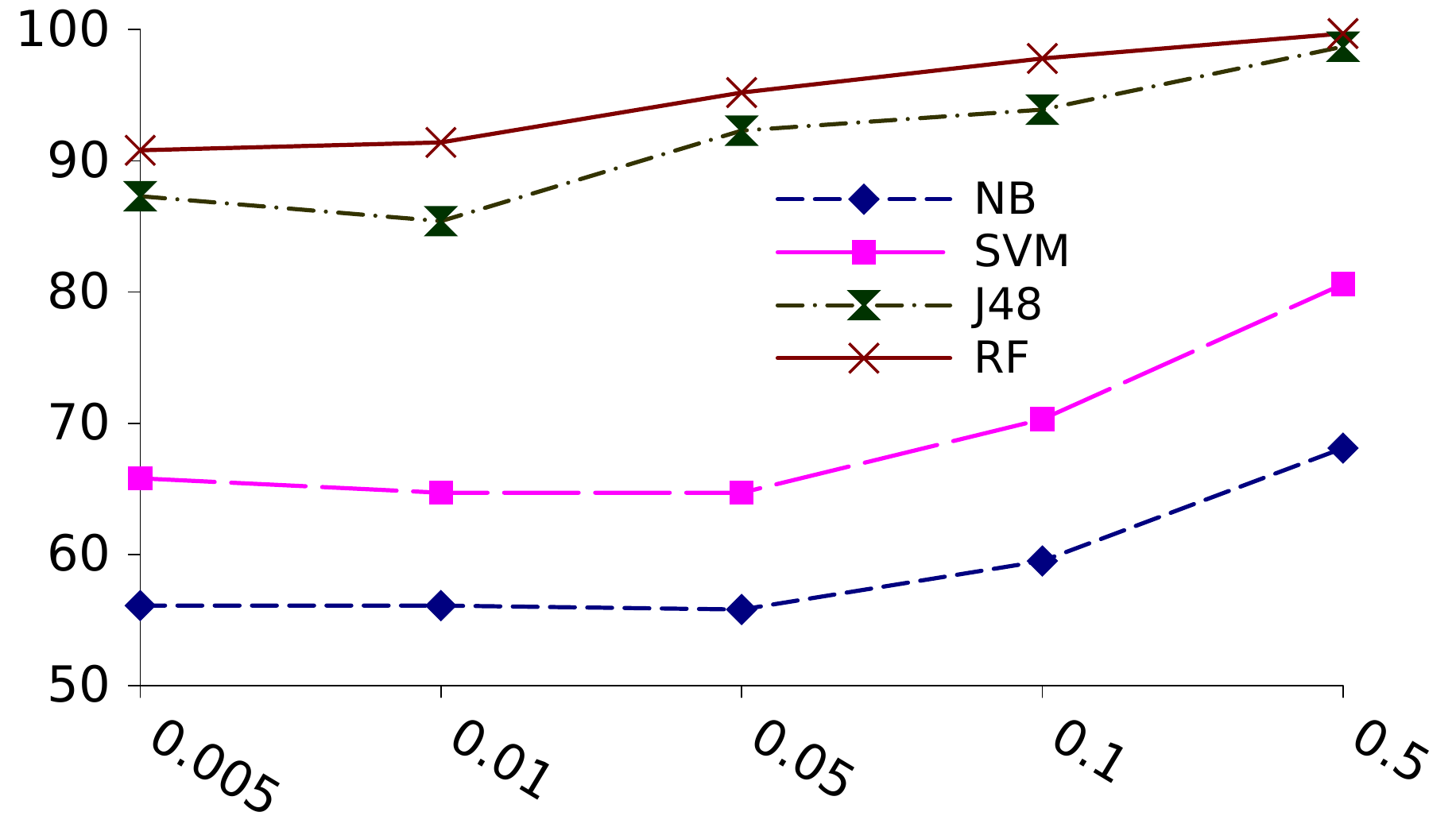}
                \caption{Accuracy of the four different classifiers over the different values of $\theta$}
                \label{fig:accuracylambda}
\end{figure}

There are several factors that may affect the performance of classification algorithms, which can also be used to explain this. These include randomness and sparsity of the actual dataset, the probability of noises and outliers, the size of the dataset, and the number of independent features -- i.e., dimensionality. In addition, many algorithms need calibration to perform well~\cite{caruana2006ECSLA}.
Focusing on our experiments, the results showed that tree-based classification approaches work best, of which Random Forest is the best classifier. This is because we experimented with a large  dataset that has a high degree of randomness and a high number of independent features. This conclusion is also supported by results from other studies~\cite{caruana2006ECSLA,Balog:2013:MCA}. Moreover, the fact that we applied the classifiers with default parameters, with no tuning, played an important role. Since we focus on the ability to treat the classifiers as a "black-box", squeezing out every bit of performance by tweaking the classifiers' parameters is beyond the scope of this work. In conclusion, we choose Random Forest as the base classifier for our framework. 

\subsection{Retrieval Effectiveness Comparison}
\label{subsec:retrievaleffectivenesscomparison}
As part of our evaluation, we performed a comparative study on the retrieval performance. We compared our approaches with the baseline methods by executing a standard retrieval model -- i.e., the \textit{BM25}, and applying the query expansion models described in the previous section -- i.e., \textit{Kullback-Leibler (KL)} divergence, in combination with both the temporal features (\textit{KL\_T}) and with the spatio-temporal features (\textit{KL\_ST}). More specifically, we used the Rocchio's framework weighting model, with both the KL divergence model to choose the expansion terms. For each query expansion run, we used the default value of $\beta$ from \cite{Araujo2008} -- i.e., $\beta=0.4$, and chose the first $n$ terms of the top-$k$ documents for the \textit{Rocchio's Beta} weighting model. The numbers of pseudo relevant documents, $k$, were set to 20, 40, 60, 80, 100, and 120, and the numbers of selected terms, $n$,  were 15, 25, 35, 45, and 55. Finally, we performed the query expansion baseline model based on the geo-temporal tag similarities as proposed by \citet{Zhang2012} -- i.e., the \ZKYC~discussed in Section~\ref{subsec:comparison-related-work}. 

\begin{table}[!ht]
\small
\tabcolsep 2pt
\small
\center
\begin{tabular}{c|c||r|c||c||c|c}
\toprule
\multicolumn{2}{c||}{} & \multicolumn{2}{c||}{\textbf{Baseline}} &\textbf{ Related method} & \multicolumn{2}{c}{\textbf{Our approaches}}\\
\midrule
 \textbf{\#Doc} & \textbf{\#Term} & \textit{BM25}	&\textbf{KL}& \textbf{KL+ZKYC\cite{Zhang2012}}	&\textbf{KL\_T}	& \textbf{KL\_ST}  \\
\midrule
	20	&15	&0.4448	&$0.4601$ & $0.4614$&$0.4752^{12}$ &$0.4816^{12}$ \\
	&25	&$0.4448$	&$0.4605$ & $0.4626$&$0.4755^{12}$	&$0.4838^{123}$ \\
	&35	&$0.4448$	&$0.4618$ & $0.4638$&$0.4761^{12}$	&$\mathbf{0.4838}^{123}$  \\
	&45	&$0.4448$	&$0.4618$	 &$0.4634$&$0.4764^{12}$	&$0.4833^{123}$  \\
	&55	&$0.4448$	&$0.4618$ &$0.4624$&$0.4761^{12}$	&$0.4838^{123}$  \\
	\midrule
40	&15	&$0.4448	$&$0.4708$	& $0.4744$ &$0.4786	^{12}$&$0.4870^{123}$ \\
	&25	&$0.4448	$&$0.4714$	 & $0.4738$&$0.4799^{12}	$&$0.4880^{123}$ \\
	&35	&$0.4448	$&$0.4705$	& $0.4734$&$0.4813^{12}	$&$0.4885^{123}$  \\
	&45	&$0.4448	$&$0.4726$	& $0.4757$&$0.4843	^{12}$&$\mathbf{0.4918}^{123}$  \\
	&55	&$0.4448$&$0.4717$	& $0.4745$&$0.4827^{12}$	&$0.4913^{123}$  \\
	\midrule
60	&15	&$0.4448	$&$0.4665$	& $0.4674$&$0.4816^{12}$	&$0.4848^{12}$  \\
	&25	&$0.4448	$&$0.4685$	& $0.4696$ &$0.4818^{12}$ 	&$0.4909^{123}$ \\
	&35	&$0.4448	$&$0.4704$	& $0.4733$&$0.4856^{12}	$	&$0.4951^{123}$  \\
	&45	&$0.4448	$&$0.4712$	& $0.4721$&$0.4877^{12}	$	&$0.4957^{123}$  \\
	&55	&$0.4448$&$0.4703$	& $0.4731$&$0.4867^{12}	$	&$\mathbf{0.4957}^{123}$  \\
	\midrule
80	&15	&$0.4448$	&$0.4697$	&$ 0.4706$&$0.4803^{12}$	&$0.4894^{123}$ \\
	&25	&$0.4448$	&$0.4699	$&$ 0.4711$	&$0.4847	^{12}$	&$0.4935^{123}$ \\
	&35	&$0.4448$	&$0.4718	$&$ 0.4733$	&$0.4862	^{12}$	&$0.4951^{123}$ \\
	&45	&$0.4448	$	&$0.4712	$&$ 0.4731$	&$0.4884^{12}$	&$0.4979^{123}$ \\
	&55	&$0.4448	$	&$0.4719	$&$ 0.4741$	&$0.4890	^{12}$	&$\textbf{0.5001}^{123}$ \\
	\midrule
100	&15	&$0.4448	$&$0.4613$	&$0.4621$&$0.4701^{12}$	&$0.4727^{12}$ \\
	&25	&$0.4448	$&$0.4611$	 &$0.4619$&$0.4755^{12}	$	&$0.4802^{12}$\\
	&35	&$0.4448	$&$0.4634$	 &$0.4642$&$0.4781	^{12}$	&$0.4849^{123}$\\
	&45	&$0.4448	$&$0.4613$	&$0.4621$&$0.4803^{12}$	&$0.4879^{123}$ \\
	&55	&$0.4448	$&$0.4621$	 &$0.4631$&$0.4820^{12}	$	&$\textbf{0.4891}^{123}$\\
	\midrule
120	&15	&$0.4448	$&$0.4592$	&$0.4601$&$0.4681	^{12}$&$0.4709^{12}$ \\
	&25	&$0.4448	$&$0.4589$	&$0.4610$&$0.4769	^{12}$&$0.4814^{12}$ \\
	&35	&$0.4448	$&$0.4606$	&$0.4625$&$0.4774	^{12}$&$0.4870^{123}$ \\
	&45	&$0.4448	$&$0.4589$	&$0.4606$&$0.4829	^{12}$&$0.4899^{123}$ \\
	&55	&$0.4448	$&$0.4595$	&$0.4606$&$0.4845	^{12}$&$\textbf{0.4914}^{123}$  \\

\bottomrule
\end{tabular}
\caption{MAP Comparison between baseline QE (\textit{KL}) and QE with classifier learned with baseline+temporal features (\textit{KL\_T}) and baseline+temporal+spatial features (\textit{KL\_ST}). The best scores within each row and each group are type-set boldface. The numbers 1,2,3 in the superscript in the table indicates statistical significance improvements with respect to \textit{KL}, \textit{KL+ZKYC}\cite{Zhang2012}, \textit{KL\_T}, respectively. }
\label{tab:comparison-average-prec}
\end{table} 
Table~\ref{tab:comparison-average-prec} lists the results from our experiments. As shown, the baseline query expansion method is better than the baseline \textit{BM25} in all of our tests, with the best MAP improvement of $6.2\%$. We can also see that ranking the feedback tags using \textit{KL} and \ZKYC~to select query expansion terms does not significantly improve  the effectiveness of the ranking score of \textit{KL}. This is mainly because  \ZKYC~captures the feature of a tag distribution on a fixed geo-temporal scale,  due to the size of the geo-temporal bin.

Overall, both our proposed query expansion methods outperform both of the baseline methods, for all the combinations of number of documents and number of terms. For \textit{KL\_T}, the maximum MAP improvement is $9.7\%$, while for \textit{KL\_ST}, the improvement is $12.4\%$. Moreover, studying the MAP values, both \textit{KL\_T} and \textit{KL\_ST} outperform \ZKYC. As discussed in Section~\ref{subsec:comparison-related-work}, an important difference between our method and  \ZKYC~ is the property of geo-temporal attractiveness of terms with respect to scales. Recall that with \ZKYC, the geo-temporal features are extracted using a fixed scale. In contrast, our methods allow extracting the features at different scales, and take spatial attractiveness into account (see Section~\ref{sec:preliminary}). Because the concentration of pictures normally vary both in time and space, considering spatial attractiveness and scales is important. The above results further confirm this importance.  In conclusion, the ability to capture the geo-temporal attractivenesses of terms at different geo-temporal scales leads to improved retrieval performance.

\begin{figure}[!ht]
	\centering

				\includegraphics[height=3.5cm]{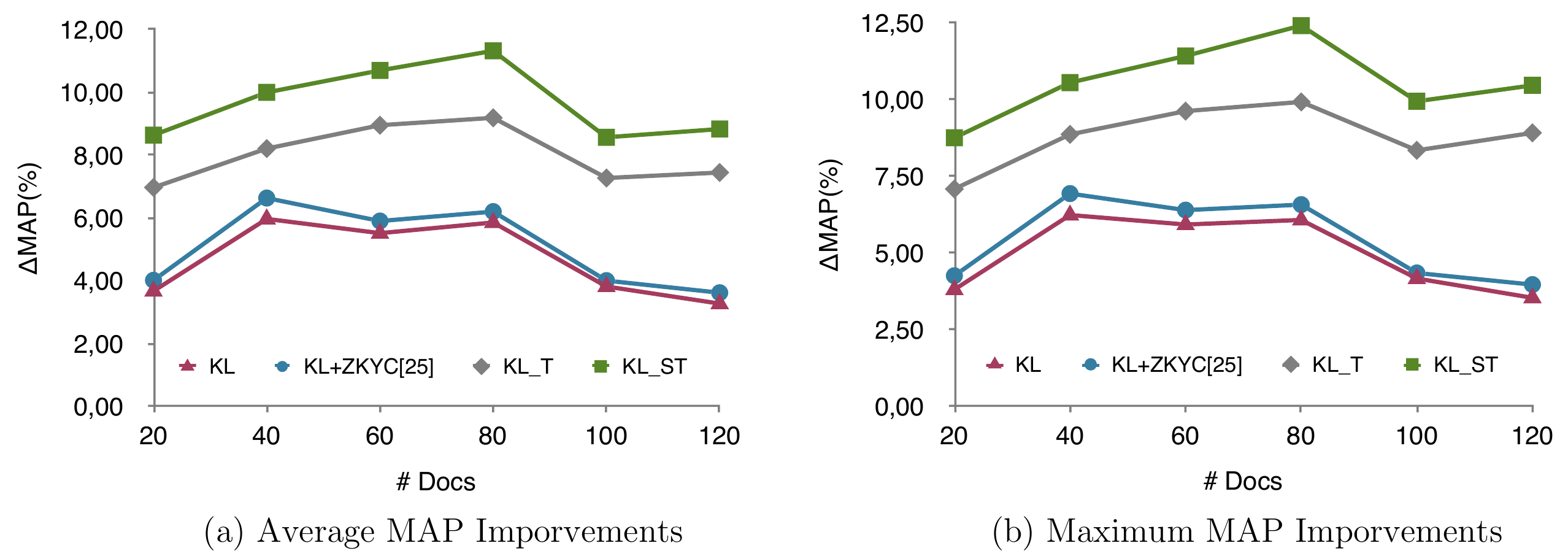}
                \caption{Comparison of MAP  improvements as function of feedback documents}
                \label{fig:max-gain-ap}
\end{figure} 
In Figure~\ref{fig:max-gain-ap}, we summarize the improvements of MAP compared to \textit{BM25}, while varying the numbers of feedback documents ($k$).  Specifically, in  Figure~\ref{fig:max-gain-ap}a, for each method, we first take the average MAP values for different numbers of terms. Then, we plot the values as function of the number of feedback documents. In Figure~\ref{fig:max-gain-ap}b, on the other hand, we plot the best MAP values for each method by only taking into account the numbers of feedback documents, independent of the number of terms. As we can observe in both graphs, all the four query expansion methods have similar trends; that is, the search process using each method gains benefits from the query expansion until reaching a specific number of documents -- i.e., a breakpoint, and thereafter this benefit decreases. However,  the breakpoint for both of our two approaches (\textit{KL\_T} and \textit{KL\_ST}) is much higher than with both the baseline \textit{KL} and \ZKYC~-- i.e., 80 versus 40.  The reason for this is that with baseline \textit{KL}, the set of candidate expansion terms are explored by considering only document features, which seems to be too restrictive. Moreover, \ZKYC~does not consider spatial attractiveness and variation in scales. 

%
	

\subsection{Analysis of the Features}
\label{subsec:featureanalysis}
In this section, we analyse the effectiveness the temporal and spatial features we have used to learn the classifiers for selection of the expansion terms. The question we want answered is: Do the features we have proposed in this paper contribute to improve the classification accuracy, and which features work best? To ensure comprehensiveness,  we perform our analyses using three different widely-used correlation-based feature evaluation methods. More specifically, we use \textit{Information Gain} (\textit{IG})~\cite{Yang1997}, \textit{Gain Ratio} (\textit{GR})~\cite{Forman2003} and \textit{Symmetrical Uncertainly} (\textit{SU})~\cite{yu2003}. Information Gain is given by 
	$IG({C},{F}) = \mathcal{H}({C}) - \mathcal{H}(C | F),$
where $\mathcal{H}(\textit{C})$ is the entropy of a class $C$ and $\mathcal{H}(C | F)$ is the entropy of the class, given a feature $F$. 
Gain Ratio is the direct extension of Information Gain, which is
	$GR({C},{F}) = {IG({C},{F})}/{\mathcal{H}({C})}.$
%
Symmetrical Uncertainly (SU) evaluates the goodness of a subset of features $F$ by comparing its symmetrical uncertainty with another subset of features~\cite{yu2003}. Let ${F_{Sub_1}} \subset F$ and ${F_{Sub_2}} \subset F$ such two subsets. Then,
$SU(F_{Sub_1},F_{Sub_2}) = {IG({F_{Sub_1}},{F_{Sub_2}})}/({\mathcal{H}({F_{Sub_1}})+\mathcal{H}(F_{Sub_2})})$.
As before, we use  Weka to implement of the feature selection methods. 

Table~\ref{tab:featureselection-ig}, \ref{tab:featureselection-gr} and \ref{tab:featureselection-su} report the IG, GR and SU scores, respectively, for the features we used in our classification of good and bad expansion terms. They show which features are the best using the {\em baseline and temporal features} compared with applying {\em baseline, temporal and spatial features}.
\begin{table}[!ht]
\vspace*{-0.5\baselineskip} 
\tabcolsep 2pt
\small
\center
\begin{tabular}{lr|lr}
\toprule
\multicolumn{2}{c}{\textbf{Baseline+Temporal}} & \multicolumn{2}{c}{\textbf{Baseline+Spatial+Temporal}}\\
\midrule
Feature & IG Score & Feature & IG Score \\
\midrule
$coOccSingle_{Whole}$ & 0.104 & $AC2	$ & 0.204\\
$CC$				 & 0.066 & $RDPA2_{Second}[1]$	 & 0.106\\
$KURT12$			 & 0.065 & $RDMD12[2]$	 & 0.097\\
$AC12$				 & 0.046 & $RDPA12[3]$ & 0.088\\
$DF3_{Feedback}$	 & 0.035 & $RDMD12[1]$	 & 0.086\\
$coOccSingle_{Feedback}$	 & 0.034 & $RDMD12[3]$ & 0.086\\
$coOccPair_{Feedback}$ & 0.031 & $RDPA12_{First}[3]	$ & 0.080\\
$DF0_{Feedback}$	 & 0.029 & $RDPA12[1]$ & 0.074\\
$DF1_{Feedback}$	 & 0.025 & $DF3_{Feedback}$	 & 0.064\\
$DF2_{Feedback}$	 & 0.025 & $RDMD12_{Second}[1]$	 & 0.063\\
$DF2_{Whole}$		 & 0.021 & $RDPA12_{First}[1]	$ & 0.062\\
$DF0_{Whole}$		 & 0.021 & $RDPA12[2]$	 & 0.062\\
$DF1_{Whole}$		 & 0.021 & $RDMD12[4]$	 & 0.056\\
$DF3_{Whole}$		 & 0.021 & $KURT2$	 & 0.048\\
$coOccPair_{Whole}$	 & 0.021 & $coOccSingle_{Whole}$	 & 0.048\\
$KURT1$				 & 0.000 &  & \\
$AC1$				 & 0.000 &  & \\
\bottomrule
\end{tabular}
\caption{Comparison of the feature quality based on Information Gain. RDMDs are the features related to the relative discrete
maximum distance feature vectors. RDPAs are the features related to the relative discrete
positive area vectors. "L" means that we use a cross-L (or cross-D) function. "\textit{First}" and "\textit{Second}" stand for first and second order feature, respectively. "[number]" denotes the number, $k$, of intervals $h_k$ used to compute the $L$ (or $D$) function (see Section~\ref{sec:proposedframework}).}
\label{tab:featureselection-ig}
\end{table}

\begin{table}[!ht]
\vspace*{-0.5\baselineskip} 
\tabcolsep 2pt
\small
\center
\begin{tabular}{lr|lr}
\toprule
\multicolumn{2}{c}{\textbf{Baseline+Temporal}} & \multicolumn{2}{c}{\textbf{Baseline+Spatial+Temporal}}\\
\midrule
Feature & RG Score & Feature & RG Score \\
\midrule
$KURT12$			 & 0.104 & $RDMD12_{Second}[4]$	 & 0.157\\
$AC12$				 & 0.086 & $RDPA12_{First}[4]$	 & 0.116\\
$coOccSingle_{Feedback}$	 & 0.079 & $RDPA12_{Second}[3]$	 & 0.111\\
$coOccSingle_{Whole}$ & 0.056 & CC	 & 0.109\\
$CC$				 & 0.054 & $RDMD12_{First}[4]$	 & 0.109\\
$DF0_{Feedback}$	 & 0.040 & $RDPA12[3]$	 & 0.107\\
$DF3_{Feedback}$	 & 0.040 & $RDPA12_{Second}[4]$	 & 0.107\\
$DF1_{Feedback}$	 & 0.036 & $RDMD12[3]$	 & 0.105\\
$DF2_{Feedback}$	 & 0.036 & $RDMDL12_{Second}[2]$	 & 0.098\\
$coOccPair_{Feedback}$ & 0.031 & $RDMD12_{Second}[2]$	 & 0.093\\
$coOccPair_{Whole}$ & 0.025 & $RDMD12[2]$	 & 0.093\\
$DF3_{Whole}				$ & 0.025 & $RDMDL12_{Second}[2]$	 & 0.088\\
$DF2_{Whole}$		 & 0.025 & $RDMD12[4]$	 & 0.085\\
$DF0_{Whole}$		 & 0.025 & $RDMDL12[4]$	 & 0.078\\
$DF1_{Whole}$		 & 0.025 & $RDPAL12[2]$	 & 0.078\\
$KURT1$			 & 0.000 &  & \\
$AC1$				 & 0.000 &  & \\
\bottomrule
\end{tabular}
\caption{Comparison of the feature quality based on Gain Ration. RDMDs are the features related to the relative discrete
maximum distance feature vectors. RDPAs are the features related to the relative discrete
positive area vectors. "L" means that we use a cross-L (or cross-D) function. "\textit{First}" and "\textit{Second}" stand for first and second order feature, respectively. "[number]" denotes the number, $k$, of intervals $h_k$ used to compute the $L$ (or $D$) function (see Section~\ref{sec:proposedframework}).}
\label{tab:featureselection-gr}
\end{table}

\begin{table}[!ht]
\vspace*{-0.5\baselineskip} 
\tabcolsep 2pt
\small
\center
\begin{tabular}{lr|lr}
\toprule
\multicolumn{2}{c}{\textbf{Baseline+Temporal}} & \multicolumn{2}{c}{\textbf{Baseline+Spatial+Temporal}}\\
\midrule
Feature & SU Score & Feature & SU Score \\
\midrule
$KURT12$	 & 0.080 & $AC2$	 & 0.107\\
$coOccSingle_{Whole}$	 & 0.073 & $RDPA12[3]$	 & 0.096\\
$AC12$	 & 0.060 & $RDMD12[2]$ & 0.095\\
$CC$	 & 0.059 & $RDMD12[3]$ & 0.094\\
$coOccSingle_{Feedback}$	 & 0.048 & $RDMD12[4]$ & 0.067\\
$DF3_{Feedback}$	 & 0.037 & $RDPA12_{First}[3]$	 & 0.064\\
$DF0_{Feedback}$	 & 0.034 & $RDPA12[2]$ & 0.063\\
$coOccPair_{Feedback}$	 & 0.031 & $RDPA12_{First}[1]$	 & 0.062\\
$DF1_{Feedback}$	 & 0.029 & $RDMD12_{Second}[1]$ & 0.057\\
$DF2_{Feedback}$	 & 0.029 & $RDPA2_{Second}[1]$ & 0.057\\
$DF2_{Whole}$	 & 0.023 & $RDPA12[4]$	 & 0.054\\
$DF0_{Whole}$	 & 0.023 & $RDPAL12_{First}[2]$  & 0.054\\
$DF1_{Whole}$	 & 0.023 & $DF3_{Feedback}	$	 & 0.053\\
$DF3_{Whole}$	 & 0.023 & $RDMD12[1]$	 & 0.052\\
$coOccPair_{Whole}$	 & 0.023 & $RDMD12_{First}[3]$	 & 0.051\\
$KURT1$	 & 0.000 &  & \\
$AC1$	 & 0.000 &  & \\
\bottomrule
\end{tabular}
\caption{Comparison of the feature quality based on Symmetrical Uncertainly. RDMDs are the features related to the relative discrete
maximum distance feature vectors. RDPAs are the features related to the relative discrete
positive area vectors. "L" means that we use a cross-L (or cross-D) function. "\textit{First}" and "\textit{Second}" stand for first and second order feature, respectively. "[$k$]" denotes the number, $k$, of intervals $h_k$ used to compute the $L$ (or $D$) function (see Section~\ref{sec:proposedframework}).}
\label{tab:featureselection-su}
\end{table}

Focusing on the baseline and temporal features, these results show that with all the three feature selection methods -- i.e., IG, GR and SU, none of the features related to the temporal autocorrelation (\textit{AC1}) and kurtosis (\textit{KURT1}) have any impact on the classification. This means that the information about peaks in the temporal distribution of candidate expansion terms does not seem to have any effects on determining good candidate expansion terms. However, the temporal correlation between the distribution of documents annotated with a candidate expansion term and of those annotated with a term from the initial query -- i.e., \textit{AC12}, \textit{KURT12} and \textit{AC12}, seem important, as their scores are within the top-5 highest scores. Similar observation can be made on the cross-correlation -- i.e., \textit{CC}, between the time series of a candidate expansion term and a query term. 

Focusing on our set of features -- i.e, the {\em baseline, temporal and spatial features}, on the other hand, 
our observation is that with all the three feature selection methods, the most important features are those related to the vectors $\vector{D}_{\textit{Max}}(\Q,e)$ (called $RDMDL12$ in Table~\ref{tab:featureselection-ig}, \ref{tab:featureselection-gr} and \ref{tab:featureselection-su}), $\vector{D}_{\textit{Max}}(\Q+e)$ (or $RDMD12$), $\vector{D}_{\textit{Sum}}(\Q,e)$ (called $RDPAL12$ in Table~\ref{tab:featureselection-ig}, \ref{tab:featureselection-gr} and \ref{tab:featureselection-su}) and $\vector{D}_{\textit{Sum}}(\Q+e)$ (or $RDPAL12$). This means that the features related to the spatial distributions of the documents annotated with both the candidate expansion terms and the query terms, and the spatial correlation between the two tag point patterns have a strong impact on the classification results. As a conclusion, our analysis confirms the importance of using the spatial correlations between a candidate expansion term and a query term as features for classification of good and bad candidate expansion terms. 
\section{Conclusion}
\label{sec:conclusion}
In this work, we have developed a new approach to effectively retrieve event-based images from typical media sharing applications, such as Flickr. To achieve this, we have developed a new method using a new set of spatial features extracted from image tags to capture the characteristics of the spatial distributions of such tags. This has included applying rigorous statistical exploratory analysis of spatial point patterns to extract the geo-spatial features. As we have shown in this paper, with these features, we have been able to both summarize the spatial characteristics of the spatial distribution of a single term, and identify the similarity between the spatial profiles of two terms. Further, aiming at improving the retrieval performance, we have investigated the gain of combining our geo-spatial features with a set of temporal features from the current state-of-the-art approaches within information retrieval. In addition, we have studied the usefulness of our method by applying our features in a machine-learning-based query expansion framework. More specifically, we have used our spatial and temporal features to select of the best candidate terms for the query expansion process. The originality of this work lies in the way we extract these features and how we use them to choose the best expansion terms. Our experiments and extensive analyses, including comparison against the  baseline methods and existing work, have demonstrated the effectiveness of our approach. These have particularly shown the importance of our proposed spatial features and the feasibility of our approach.

Nevertheless, there are interesting aspects of this work that we have left for further investigation. First, to further explore the usefulness of our spatial features in more general information retrieval settings, we currently study applying our approach on other resources than pictures. Second, in this paper, we have focused on selecting candidate expansion terms as a binary classification problem. As part of making our approach even more generic, we are investigating performing unsupervised selection of  expansion terms based on their associated temporal and geo-spatial characteristics. Third, we are exploring the combination of this approach with Open Linked Data usage, such as DBPedia, to further improve the choice of best expansion term candidates.

\section*{Acknowledgement}
We acknowledge the anonymous reviewers’ comments, which have been invaluable in improving the quality of this manuscript. This work is supported by the Research Council of Norway, VERDIKT program grant number 176858.

%
\small
\bibliographystyle{elsarticle-num-names}
\setlength{\bibsep}{0.5\parsep} 
\bibliography{ruocco_ramampiaroIPM}  
\end{document}